\documentclass[a4paper,11pt]{article}
\usepackage{jheppub}

\usepackage[utf8]{inputenc}
\usepackage{amsmath}
\usepackage{amssymb}
\usepackage{mathrsfs}
\usepackage{graphicx}
\usepackage{bbold}
\usepackage{comment}
\usepackage[dvipsnames]{xcolor}
\usepackage{braket}
\usepackage{bm,bbm}
\usepackage{physics}
\usepackage{tikz}
\usepackage{mathtools}

\newcommand\beq{ \begin{eqnarray} }
\newcommand\eeq{ \end{eqnarray} }

\usepackage{bbm}
\usepackage{physics}
\usepackage{simplewick}

\preprint{YITP-26-46, RIKEN-iTHEMS-Report-26}

\title{Hadron spectra of finite-density QC$_2$D}

\author[a,b]{Kei Iida,}
\author[c,d]{Etsuko Itou,}
\author[d,e]{Kotaro~Murakami,}
\author[f]{and Daiki~Suenaga}
\affiliation[a]{Department of Liberal Arts, The Open University of Japan, Chiba 261-8586, Japan }

\affiliation[b]
{RIKEN Nishina Center, RIKEN, 2-1 Hirosawa, Wako, Saitama 351-0198, Japan}

\affiliation[c]{Yukawa Institute for Theoretical Physics, Kyoto University, Kitashirakawa Oiwakecho, Sakyo-ku, Kyoto 606-8502, Japan }
\affiliation[d]{RIKEN Center for Interdisciplinary Theoretical and Mathematical Sciences (iTHEMS), RIKEN,
2-1 Hirosawa, Wako, Saitama 351-0198 Japan }

\affiliation[e]{Department of Physics, Institute of Science Tokyo, 2-12-1 Ookayama, Megro, Tokyo 152-8551, Japan  }

\affiliation[f]{Kobayashi-Maskawa Institute for the Origin of Particles and the Universe, Nagoya University, Nagoya, 464-8602, Japan}

\emailAdd{k-iida@ouj.ac.jp}
\emailAdd{itou@yukawa.kyoto-u.ac.jp}
\emailAdd{kotaro.murakami@yukawa.kyoto-u.ac.jp}
\emailAdd{suenaga.daiki.j1@f.mail.nagoya-u.ac.jp}

\abstract{
We investigate the chemical-potential dependence of hadron spectra in two-color QCD using first-principles lattice simulations.
We compute two-point correlation functions for all allowed hadronic operators by newly including the contributions from disconnected diagrams, and extract the corresponding effective masses.
In the meson sector, the mass hierarchy in the hadronic phase (normal vacuum) is found to be $m_\pi \lesssim m_{\eta}  < m_\sigma \mathrm{(noisy)} < m_\rho \sim m_\omega  \ll m_{a_1}$, which is similar to that in three-color QCD.
In the superfluid phase, this hierarchy is modified, and with increasing density it changes to $m_\sigma \mathrm{(noisy)}  < m_{a_1} < m_\rho <  m_\pi \sim m_{\eta} \mathrm{(noisy)} \ll m_{\omega} \mathrm{(noisy)}$.
In the diquark sector, the ordering remains as $m_{NG} \lesssim m_{I=0, S} < m_{I=1, AV} < m_{I=0, PS} \lesssim m_{I=0, V}$ in both phases, and the Nambu–Goldstone mode associated with spontaneous breaking of $U(1)_B$ is confirmed to be nearly massless.
Furthermore, by comparing correlators for chiral partners, we find indications of chiral symmetry restoration at high density.
}

\begin{document}
\maketitle
\flushbottom
\section{Introduction}
\label{sec:intro}

The equilibrium phase diagram of dense matter has yet to be theoretically determined, due mainly to the sign problem that occurs in Monte Carlo simulations of finite-density QCD, but has been increasingly constrained by neutron-star observations, in particular since gravitational waves from neutron star mergers were detected~\cite{MUSES:2023hyz}. 
Recent constraints on the equation of state suggest that the speed of sound in neutron-star matter may develop a peak above normal nuclear density.
At low temperatures and high densities, chiral symmetry is expected to show a tendency toward restoration, accompanied by a reduction of the chiral condensate ($\langle \bar{q}q\rangle$).
In addition, diquark condensation may occur, leading to color-superconducting phases in three-color QCD. 
Such high-density phases may be relevant to the physics of neutron-star interiors.

A natural question is how the properties of hadronic excitations are modified in such phases, which are distinct from the normal QCD vacuum.
Since hadron spectra are sensitive to the underlying condensates and symmetry-breaking pattern, it is natural to expect that they are modified in medium.
Thus far, many efforts have been made to elucidate the connection between in-medium hadron properties and chiral symmetry restoration.
One of the earliest and most influential ideas is the Brown–Rho scaling~\cite{Brown:1991kk}, which proposed that in-medium masses of a nucleon as well as scalar ($\sigma$) and vector ($\rho$, $\omega$) mesons scale with the in-medium pion decay constant $f_\pi^*$.  
According to this scaling, these masses decrease as chiral symmetry is partially restored with increasing medium density.  
In other chiral models, various modifications of hadron properties in medium have also been discussed (see, e.g., Refs.~\cite{Hatsuda:1994pi,Rapp:1999ej} for reviews and references therein). 
QCD sum-rule analyses by Hatsuda and Lee suggested a downward shift of vector-meson masses in medium~\cite{Hatsuda:1991ez, Hatsuda:1995dy, Hatsuda:1996az}. 
Later, it was argued that such mass shifts are difficult to disentangle from the broadening of the corresponding resonances~\cite{Klingl:1997kf, Rapp:1999ej}; moreover, other QCD sum-rule analyses claimed that the $\rho$ and $\omega$ meson masses do not necessarily decrease in medium, while that of the $\phi$ meson may decrease~\cite{Zschocke:2002mn,Steinmueller:2006id}.  
Furthermore, the anomalous breaking of $U(1)_A$, namely, instanton effects, is expected to be suppressed at high density~\cite{Schafer:1999fe}; it has been suggested that a similar mechanism also operates in two-color QCD (QC$_2$D), leading to a reduction of the $\eta^{\prime}$ mass (or the $\eta$ mass in the two-flavor case) in the superfluid phase~\cite{Kanazawa:2009ks, Kawaguchi:2023olk}. 
Theoretical discussions of vector-meson modifications have motivated experimental searches in nuclear matter, such as the KEK-PS E325 and J-PARC E16 experiments~\cite{Philipp2014, Philipp2025,Ichikawa:2025, Aoki:2021cqa, JPARC-E-016:2023syy}. 
Recently, a possible mass reduction of the $\eta'$ meson in medium has also been suggested by the $\eta'$-mesic nuclei experiment at GSI~\cite{e-PRiME:2025fer}.

On the ohter hand, the most established first-principles approach to QCD at the zero quark-chemical potential, $\mu=0$, (normal QCD vacuum) is Monte Carlo simulation of lattice QCD. 
Nowadays, lattice QCD calculations can be performed with quark masses close to the physical point, and many hadron masses are reproduced consistently with experimental results using only a few input parameters, such as the gauge coupling and bare quark masses (e.g., Ref.~\cite{Aoyama:2024cko}). 
However, once one attempts to explore the finite-density regime, the notorious sign problem prevents direct Monte Carlo simulations.

\begin{figure}[htbp]
\centering
\includegraphics[width=.45\textwidth]{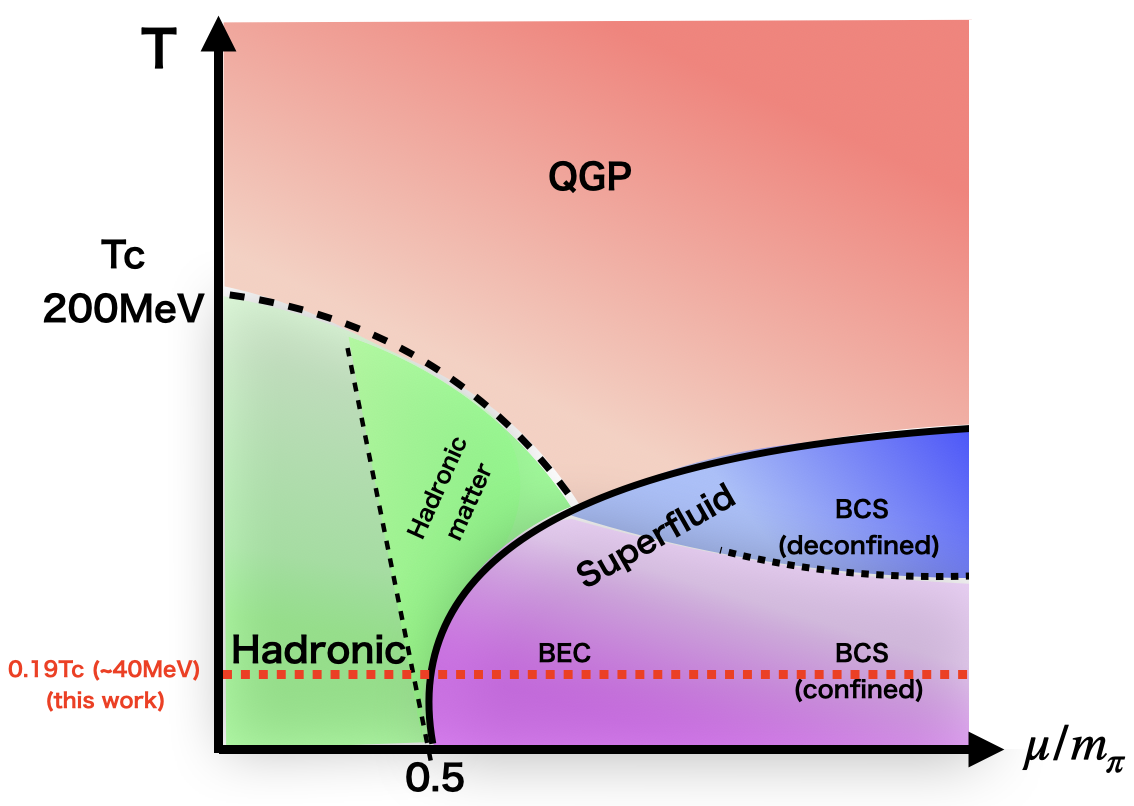}
\caption{The QC$_2$D phase diagram. \label{fig:phase-diagram}}
\end{figure}
One possible way to circumvent this difficulty is to study two-color QCD instead of three-color QCD. 
This approach is useful because it avoids the sign problem at finite density while at $\mu=0$ it still shares fundamental QCD phenomena such as confinement and chiral symmetry breaking. Furthermore, the equilibrium phase diagram of two-color QCD, as shown in Fig.~\ref{fig:phase-diagram}, has been extensively studied by systematically calculating the color-singlet diquark condensate ($\langle qq\rangle$), the Polyakov loop ($\langle |L|\rangle$), and the (net) quark number density ($\langle n_q\rangle$) as functions of the quark chemical potential $\mu$ and the temperature $T$~\cite{Skullerud:2003yc, Allton:2003vx, Hands:2006ec, Hands:2006ve,   Hands:2007uc, Hands:2011ye, Hands:2011hd,  Cotter:2012mb, Hands:2012fs, Boz:2015ppa, Boz:2019enj, Hands:2024nkx, Braguta:2016cpw, Astrakhantsev:2018uzd,Astrakhantsev:2020tdl, Begun:2022bxj,Buividovich:2020dks, Buividovich:2020gnl, Buividovich:2021fsa, Iida:2019rah, Iida:2020emi,Iida:2022hyy, Ishiguro:2021yxr, Iida:2024irv} (see also a review paper~\cite{Itou:2025vcy}). 
At sufficiently high temperatures, a quark-gluon plasma with a clearly nonzero $\langle |L|\rangle$ appears, while at low temperatures, several distinct phases occur as $\mu$ increases. 
Around $\mu\simeq m_\pi/2$, the hadronic phase, where $\langle qq\rangle$, $\langle |L|\rangle$, and $\langle n_q\rangle$ are vanishingly small, undergoes a phase transition to a Bose-Einstein condensation (BEC)-type superfluid phase, where nonzero $\langle qq\rangle$ and $\langle n_q\rangle$ behave as predicted by ChPT. 
With a further increase in $\mu$, the BEC phase does not undergo any phase transition, but crosses over into a BCS-type superfluid phase, where the $\mu$-dependence of $\langle qq\rangle$ is consistent with weak-coupling expectations.
In our previous works~\cite{Iida:2019rah, Iida:2024irv}, we determined the crossover scale between these two regimes from the behavior of $\langle n_q\rangle$. Following Ref.~\cite{Hands:2006ve}, we identify the region where $\langle n_q\rangle$ is consistent with the free-quark behavior,$n_q^{tree}$, as the BCS regime, and define the lower bound of this region in $\mu$ as the BEC-BCS crossover scale.
Moreover, first-principles lattice calculations in QC$_2$D have shown that the speed of sound exceeds the conformal value ($c_{\rm s}^2/c^2=1/3$). 
This qualitative behavior resembles that inferred from recent neutron-star observations.

Hadron spectroscopy in dense QC$_2$D has been investigated in pioneering simulations by Muroya \textit{et al.}~\cite{Muroya:2002ry}, who found that the pion mass shows little dependence on $\mu$, while the $\rho$ meson mass decreases in high density regimes. 
Subsequently, Hands \textit{et al.}~\cite{Hands:2007uc} carried out more systematic studies of various hadronic channels and showed, in particular, that a light Nambu-Goldstone (NG) mode appears in the superfluid phase. 
Related studies using staggered fermions have also reported the NG spectrum in the superfluid phase~\cite{Wilhelm:2019fvp}.
In those earlier studies, however, the contributions from disconnected diagrams were not fully incorporated. 
Hadronic masses in dense QC$_2$D have also been studied using effective field theories such as ChPT~\cite{Kogut:1999iv,Kogut:2000ek} and the sigma model~\cite{Suenaga:2022uqn,Suenaga:2025sln}. 
Nevertheless, from a first-principles point of view, determining how hadron masses change with the chemical potential $\mu$ remains an important and nontrivial problem. 
In the present work, we perform a comprehensive lattice analysis of hadron spectra including the contributions from disconnected diagrams.

The remainder of this paper is organized as follows. 
Section~\ref{sec:hadron-QC2D} reviews the properties of hadrons in two-flavor QC$_2$D, including the flavor symmetry, theoretical expectations of hadron masses at finite density, and the relation between chiral symmetry and correlation functions for chiral partners. 
After introducing the lattice setup and simulation details in Sec.~\ref{sec:setup}, we present in Sec.~\ref{sec:results} the numerical results for the hadron spectra and summarize their chemical-potential dependence. 
We then turn in Sec.~\ref{sec:result-chiral-sym} to the correlation functions for chiral partners and discuss their implications for chiral symmetry restoration. 
Finally, Sec.~\ref{sec:summary} is devoted to the summary and future prospects.

\section{Hadrons in two-flavor QC$_2$D}\label{sec:hadron-QC2D}
Let us start with the definitions of hadronic operators in $N_f = 2$ QC$_2$D. Here we discuss the underlying flavor symmetry and its breaking, as well as the possible hadronic channels.

It is well known that QC$_2$D possesses an extended flavor symmetry, namely, the Pauli–G\"{u}rsey symmetry, in the limit $m = \mu = 0$ since quarks belong to a pseudo-real representation.  We review how this symmetry is broken once 
$m$ and $\mu$ are turned on~\cite{Kogut:2000ek, Kogut:2001na, Kanazawa:2009ks}, and discuss allowed hadronic operators in $N_f=2$ QC$_2$D in Sec.~\ref{sec:flavor-sym}.
In Sec.~\ref{sec:theoretical-prediction}, we summarize analytical predictions for the hadron spectra in finite-density regimes of two-color and three-color QCD.
In Sec.~\ref{sec:chiral-partner}, we discuss the relationship between the restoration of chiral symmetry and the two-point functions of hadrons that form chiral partners.

\subsection{Flavor symmetry and allowed hadronic operators}\label{sec:flavor-sym}
We start with the Euclidean Lagrangian of QCD,
\beq
\mathcal{L}_{QCD}=
  \tfrac{1}{4} F_{\mu\nu}^a F_{\mu\nu}^a
  + \sum_{f=1}^{N_f} \bar\psi^i_f (\gamma_\mu D^{ij}_\mu + m \delta^{ij}) \psi^j_f. 
\label{eq:QCD-action}
\eeq
In the two-flavor massless ($N_c=3$, $N_f=2$, and $m=0$) case, it is invariant under independent $U(2)$ rotations of the left- and right-handed fermions, so that its flavor symmetry is given by
$U(2)_L \times U(2)_R = SU(2)_L \times SU(2)_R \times U(1)_B \times U(1)_A$
as a decomposition. Among these groups, $U(1)_A$ is broken by quantum effects, while the non-Abelian chiral symmetry $SU(2)_L \times SU(2)_R$ is spontaneously broken to $SU(2)_V$ in the normal QCD vacuum. Thus, the massless $N_f = 2$ QCD vacuum is invariant under $SU(2)_V \times U(1)_B$.

\begin{figure}[htbp]
\centering
\includegraphics[width=.95\textwidth]{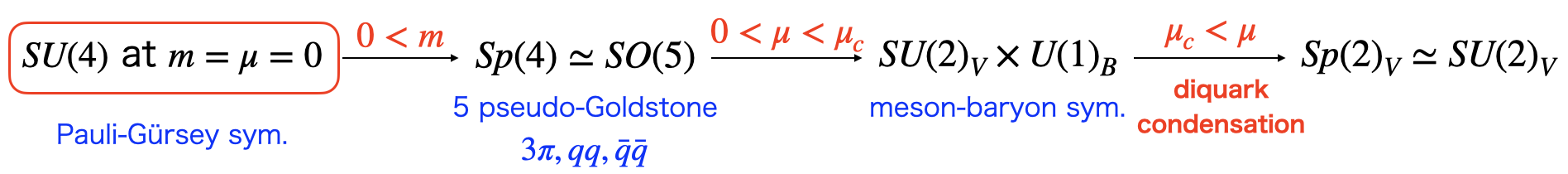}
\caption{Summary of flavor symmetry for $N_c=N_f=2$ QCD~\cite{Kogut:2000ek}. \label{fig:flavor-sym}}
\end{figure}

Now, let us consider the $N_c = N_f = 2$ case following Ref.~\cite{Kogut:2000ek}; the symmetry-breaking pattern is summarized 
in Fig.~\ref{fig:flavor-sym}.
For $N_c = N_f = 2$, the flavor symmetry is enhanced to $SU(4)$, which is known as the Pauli–G\"{u}rsey symmetry. Here $U(1)_B$ is embedded in $SU(4)$ as a diagonal generator.
As in the three-color case, $U(1)_A$ is broken by quantum effects.

If we introduce a small quark mass $m$, then the flavor symmetry is explicitly broken to $Sp(4)$~\footnote{Here we use the convention that the $N$ in $Sp(N)$ is the number of real degrees of freedom.}.
The five pseudo-NG modes corresponding to the broken generators of $SU(4)/Sp(4)$ consist of iso-triplet pions, as well as a diquark and an antidiquark; these states belong to the same multiplet under $Sp(4)$.
If we further introduce a quark chemical potential $\mu$ via
\beq
\mathcal{L}_\mu=-\mu \sum_{f=1}^{N_f} \bar\psi_f \gamma^4 \psi_f
\eeq
into the Lagrangian, 
then up to a threshold $\mu_c$ the symmetry-breaking pattern remains $Sp(4) \rightarrow SU(2)_V \times U(1)_B$. 
In other words, this is the symmetry preserved in the standard vacuum of QC$_2$D in the presence of small $m$ as long as $\mu < \mu_c$. 
When the chemical potential exceeds the threshold, $\mu > \mu_c$, the remaining $SU(2)_V \times U(1)_B$ symmetry is spontaneously broken by diquark condensation. This corresponds to the hadronic-superfluid phase transition in dense QC$_2$D.
After the $U(1)_B$ symmetry is broken, only
$Sp(2)_V \simeq SU(2)_V$
remains unbroken.
As long as $U(1)_B$ is conserved, mesons and baryons (antibaryons) belong to different sectors characterized by their baryon number.  Once $U(1)_B$ is broken, however, this separation is no longer applicable. 
In terms of mass eigenstates, they become indistinguishable if they share the same quantum numbers such as isospin ($I$), angular momentum ($J$), and parity ($P$).  This phenomenon is known as meson–baryon mixing~\cite{Kogut:2001na}.

We define the hadronic operators as
\begin{align}
M\coloneqq \bar{\psi}\Gamma \psi (x) &\quad \mathrm{meson},\\
D\coloneqq \psi^T K\Gamma \psi(x) &\quad \mathrm{baryon\,(diquark)},\\
\bar{D} \coloneqq \bar{\psi}K\Gamma \bar{\psi}^T (x) &\quad \mathrm{antibaryon\,(antidiquark)},
\end{align}
where we take
$\Gamma = \{ 1, \gamma_5, \gamma_i, i \gamma_5 \gamma_i \}$
with $i = 1,2,3$.
Following Ref.~\cite{Hands:2007uc}, we use $K=C\gamma_5 T^2$ with the charge-conjugation matrix $C=i\gamma_0 \gamma_2=\gamma_2 \gamma_4$ in the definition of the diquark and antidiquark operators. Note that $T^2$ denotes the second Pauli matrix acting on the color indices, and hence is equivalent to $-i\epsilon^{ab}$.
With this convention, mesons and baryons with the same choice of $\Gamma$ share the same $J^P$. Indeed, the hadrons with $\Gamma = \{ 1, \gamma_5, \gamma_i, i \gamma_5 \gamma_i \}$ have the corresponding quantum numbers $J^P = \{ 0^+, 0^-, 1^-, 1^+ \}$.

The diquark operators must respect Fermi statistics, which requires
$\psi_i^T (K \Gamma)\psi_j = - \psi_j^T (K \Gamma)\psi_i$,
so that only those choices of $\Gamma$ satisfying this condition are allowed.
Since the diquark state is antisymmetric in color, in the iso-triplet sector, $I=1$, which is symmetric under flavor exchange, the spinor structure $C\gamma_5 \Gamma$ must be symmetric. This condition allows only the axial-vector diquark state with $J^P = 1^+$, i.e., $\Gamma = i\gamma_5 \gamma_i$.
In contrast, in the iso-singlet sector, $I=0$, which is antisymmetric in flavor, the spinor structure must be antisymmetric. In this case, the allowed choices are
$\Gamma = \{ 1, \gamma_5, \gamma_i \}$.
In the superfluid phase where $U(1)_B$ is broken, meson–baryon mixing can occur only between each of the $\sigma$, $\eta$, $\omega$, and $a_1$ mesons and the corresponding diquark with the same $J^P$.

Furthermore, when $U(1)_B$ is spontaneously broken, both an NG and a Higgs modes appear.
In this work, we add the diquark source term,
\beq
\mathcal{L}_j = -j (  \bar{\psi}_1 K \bar{\psi}_2^T -  \psi_2^T K \psi_1 )\label{eq:diquark-source-term}
\eeq
to the Lagrangian, to explicitly fix the direction of $U(1)_B$ breaking.
The NG mode can be represented by the same operator as the diquark source term, while the Higgs mode corresponds to fluctuations in the radial direction of the order parameter.

For later use, we summarize the hadronic operators used throughout this paper in Table~\ref{table:hadron-ops}.
\begin{table}[h]
\centering
\begin{tabular}{|l|l|}
\hline\hline
Iso-singlet ($I=0$) meson 
& $M^0 = 
\left( \bar{\psi}_1 \Gamma \psi_1 
+ \bar{\psi}_2 \Gamma \psi_2 \right)/\sqrt{2}$ \\ \hline

Iso-triplet ($I=1$) meson &
$M^1 = \bar{\psi}_1 \Gamma \psi_2 , (\bar{\psi}_1 \Gamma \psi_1 -  \bar{\psi}_2 \Gamma \psi_2)/\sqrt{2}$ 
\\ \hline

Iso-singlet ($I=0$) diquark 
& $D^0 = 
\left( 
\psi_1^T K \bar{\Gamma} \psi_2 
- \psi_2^T K \bar{\Gamma} \psi_1 
\right)/\sqrt{2}$, 
\quad for $\Gamma=\{1,\gamma_5,\gamma_i\}$ \\ \hline

Iso-triplet ($I=1$) diquark 
& $D^1 = 
\left( 
\psi_1^T K \bar{\Gamma} \psi_2 
+ \psi_2^T K \bar{\Gamma} \psi_1 
\right)/\sqrt{2}$, 
\quad for $\Gamma=\{i\gamma_5\gamma_i\}$ \\ \hline

Higgs mode $(+)$, NG mode $(-)$ 
& $D^\pm =
\left(
\bar{\psi}_1 K \bar{\psi}_2^T
\pm
\psi_1^T K \psi_2
\right)$ \\ \hline\hline
\end{tabular}
\caption{Definitions of meson and diquark operators in $N_c=N_f=2$ QCD. (Here, $\bar{\Gamma}=\gamma_4 \Gamma \gamma_4$. )}
\label{table:hadron-ops}
\end{table}
We extract the hadron masses from the corresponding two-point correlation functions.
As for the antidiquark correlator, it can be computed from the backward propagation of the diquark two-point function.

\subsection{Theoretical predictions for hadron masses at finite density}\label{sec:theoretical-prediction}
In this subsection, we summarize several analytical arguments concerning hadron masses.
First, we discuss the mass shift that occurs in the hadronic phase, partly reviewing our previous work~\cite{Murakami:2023ejc}.
Next, we discuss how the mass estimation should be modified in the superfluid phase where the baryon symmetry is broken.
Finally, we review predictions for the lightest hadrons based on QCD inequalities, mainly following Ref.~\cite{Kogut:1999iv}.

\subsubsection{Mass shift in the hadronic phase}\label{sec:mass-shift}
Let us consider how hadron masses shift at $\mu < \mu_c$.
We start with the dispersion relation of a single hadron at $\mu=0$ and $T=0$, 
where the hadron with momentum $\vec{p}$ and mass $m$ has an energy $E=\sqrt{\vec{p}^2+m^2}$. 
At $\mu>0$, such a dispersion relation can be obtained from the time dependence of a (Euclidean) two-point correlation function in QCD defined by 
\begin{eqnarray}
C(T,\mu;\tau)\coloneqq\frac{1}{Z}\Tr[e^{-\frac{1}{T}(\hat{H}-\mu \hat{N})} \hat{O}(\tau) \hat{O}^{\dag}(0) ],
\end{eqnarray}
where $Z=\Tr[e^{-\frac{1}{T}(\hat{H}-\mu \hat{N})}]$ and $\hat{N}$ denote the partition function and the quark number operator conjugate to $\mu$, respectively.
The operator $\hat{O}(\tau)$ is an arbitrary Heisenberg operator at Euclidean time $\tau$. 
Here, we assume that $\hat{O}(\tau)$ carries a definite quark number, that is, $[\hat{N}, \hat{O}(\tau)] = -n_{O}\hat{O}(\tau)$.
$n_{O}$ denotes the quark number of the eigenstate; 
$n_O = 0, N_c$, and $-N_c$ for the meson, baryon, and antibaryon operators, respectively.

In the zero-temperature limit $T \to 0$, the correlation function reads 
\begin{eqnarray}
C(\mu;\tau)  \coloneqq \lim_{T\to 0}C(T,\mu;\tau) =\bra{0} \hat{O}(\tau) \hat{O}^{\dag}(0) \ket{0} =\sum_{n}|\bra{0}\hat{O}(0)\ket{n}|^{2} e^{-(E_{n}-\mu n_{O})\tau},\label{eq:mu_dep_corr}
\end{eqnarray}
where $E_{n}$ is the energy of the state 
$\ket{n}$ at $\mu=0$.  Let us now set $\hat{O}(0)$ and $\hat{O}^{\dag}(0)$ to the annihilation and creation operator of a single particle with given momentum $\vec{p}$.
Then, from the behavior of $C(\mu;\tau)$ at large $\tau$, we obtain the dispersion relation of a single hadron as 
\begin{eqnarray}
\begin{aligned}
E(\vec{p},\mu) =  \sqrt{\vec{p}^2+m^2}-\mu n_{O}.\label{eq:disp_rel}
\end{aligned}
\end{eqnarray}
One may interpret the $\mu$-dependent term as an energy shift. 
After setting $\vec{p}=\vec{0}$ and defining the effective mass as
\beq
m_{\textrm{eff.}} = m - \mu n_O,\label{eq:mass-shift}
\eeq
however, it can equivalently be interpreted as a mass shift induced by $\mu$~\footnote{The same mass-shift equation is also predicted in the hadronic phase from analyses based on ChPT (see Eq.~(80) in Ref.~\cite{Kogut:2000ek}).}. 
Note that the effective mass, which corresponds to the exponential decay rate of the two-point correlation function in the imaginary time direction, is related to the pole mass.
We also note that in the $\mu < \mu_c$ region, the meson mass is independent of $\mu$ since $n_O=0$, while the baryon and antibaryon masses are linearly shifted by $\mu$.

The discussion developed in this subsubsection is applicable only when the quark (or baryon) number remains conserved; that is, the vacuum is an eigenstate of $\hat{N}$ with zero eigenvalue ($\hat{N}|0\rangle =0$).
In lattice QCD, this property has been proven analytically only for $0 < \mu < m_\pi/2 $ in the $T \rightarrow 0$ limit in the context of dense three-color QCD~\cite{Nagata:2012ad, Nagata:2012tc, Nagata:2012mn}.
Indeed, in our lattice Monte Carlo configurations, the expectation value of the quark number operator is almost zero for each configuration ($|\Omega_i\rangle$), i.e., $\langle \Omega_i|\hat{N} |\Omega_i\rangle =0$ without taking the configuration average.

\subsubsection{Meson-baryon mixing in the superfluid phase}\label{sec:meson-baryon-mixing}
We turn to hadron masses at $\mu > \mu_c$, i.e., in the superfluid phase.
In this phase, the vacuum assumed in Eq.~\eqref{eq:mu_dep_corr} is modified, and hence
the $\mu$-dependence of the correlation function becomes more complex.
Indeed, in the hadronic phase where the $U(1)_B$ symmetry is preserved, the meson, diquark, and antidiquark belong to different baryon-number sectors and can have different low-lying masses.
In the superfluid phase where the $U(1)_B$ symmetry is broken, on the other hand, they are no longer distinguished by baryon number and can mix with each other.
Thus, if mesons and diquarks (antidiquarks) share the same quantum numbers $(I, J^P)$, their mixing has to be properly taken into account.

Then, if we write a spectral representation analogous to Eq.~\eqref{eq:mu_dep_corr} for two-point correlators in the superfluid phase, where $\ket{0}$ is no longer an eigenstate of $\hat{N}$,
the salient feature is that the Hamiltonian eigenstates $\ket{n}$ can be written schematically as
\beq
\ket{n}
=c_n \ket{\text{state created by } M } + d_n \ket{\text{state created by } D} + \bar{d}_n \ket{\text{state created by } \bar{D}},\nonumber\\
\label{eq:M-D-mixing}
\eeq
reflecting the mixing between meson-type and diquark-type operators.
In evaluating hadronic correlation functions through quark contractions, one must include not only the normal propagator connecting $\bar{\psi}$ and $\psi$ but also the anomalous propagator connecting $\psi^T$ and $\psi$; the latter can take a nonzero value in the superfluid phase (see details in Sec.~\ref{sec:had-corr}).

To obtain the mass eigenstates, one needs to compute the correlation functions $C_{MM}$, $C_{MD}$, $C_{DM}$, and $C_{DD}$ and diagonalize them, for example, by solving a generalized eigenvalue problem (GEVP).
In the present study, however, we separately compute the two-point functions of the meson-type and baryon-type operators defined in Table~\ref{table:hadron-ops} even in the superfluid phase.
If such mixing is present, these operators can couple to the same set of energy eigenstates with identical quantum numbers. At large Euclidean time, the lightest state in this common spectrum would dominate. As a result, the masses extracted from the separate correlators may appear degenerate.

\subsubsection{QCD inequality and the lightest hadrons}\label{sec:qcd-inequality}
Here, we briefly review the predictions for the lightest hadrons based on QCD inequalities for QC$_2$D~\cite{Kogut:2000ek}.
We begin with the three-color case for comparison. 

In Euclidean three-color QCD at $\mu = 0$, two-point correlation functions are positive definite. 
It is well known that the lightest hadron is the pion, i.e., an iso-triplet channel of $\bar{\psi}\gamma_5  \psi$~\footnote{In this subsubsection, we drop the flavor structure and simply denote the iso-triplet meson as $M(x)=\bar{\psi}\Gamma\psi (x)$; more precisely, it is $M^1$ in Table~\ref{table:hadron-ops}.}. 
This can be proven from the QCD inequalities as follows.
We assume the $\gamma_5$-Hermiticity of the Dirac operator,
\beq
\gamma_5 D \gamma_5 = D^\dagger,
\eeq
and no contribution from disconnected diagrams.
Then, the correlator for a generic iso-triplet meson, $M(x) = \bar{\psi}(x)\Gamma\psi(x)$, is given by
\beq
\langle M(x)M(0)\rangle 
= \Tr\, [S(x,0)\Gamma S(0,x)\Gamma],
\label{eq:generic_corr}
\eeq
where $S$ denotes the quark propagator, $S \equiv D^{-1}$.
From the Schwarz inequality, we obtain the QCD inequality,
\beq
\Tr\, [S(x,0)\Gamma S(0,x)\Gamma]
= \Tr\, [S(x,0)\Gamma\gamma_5 S^\dagger(x,0)\gamma_5\Gamma]
\le
\Tr\, [S(x,0) S^\dagger(x,0)].
\label{eq:schwarz}
\eeq
Here, $\Tr\, [S(x,0) S^\dagger(x,0)]$ corresponds to the correlation function for the meson with $\Gamma=\gamma_5$ since $\gamma_5 D \gamma_5 = D^\dagger$.
Thus, the correlation function at $\mu=0$ for any meson is positive definite, and the meson correlation function with $\Gamma=\gamma_5$ is always larger than that for all the other mesons.  This implies that the exponential decay rate of the correlation function, i.e., the hadron mass in this channel, is smaller than in any other channel, and hence the pion is the lightest hadron at $\mu=0$.

For $\mu \neq 0$ where the positivity is lost, the above inequalities in three-color QCD no longer hold.
In dense QC$_2$D, however, because of the pseudo-reality of the fundamental representation of SU(2),
\beq
K  \Delta (\mu) K  = \Delta^*(\mu),
\label{eq:pseudoreal}
\eeq
holds even at $\mu \neq 0$. Here, $\Delta (\mu)$ denotes the Dirac operator including the quark number operator.
Equation~\eqref{eq:pseudoreal} ensures that the functional measure is positive.

Then, the correlation function for the iso-singlet scalar diquark, represented by the $\psi^T_1 K \psi_2$-type operator~\footnote{More precisely, it is $D^0$ in Table~\ref{table:hadron-ops}.}, can be written as
\beq
\Tr\, [S(x,0) K
S^T(x,0)K]
=
\Tr\, [S(x,0) S^\dagger(x,0)].
\label{eq:diquark_corr}
\eeq 
In the same way as in the $\mu=0$ case, one can show that the iso-singlet scalar diquark is the lightest among the diquark operators of the form $\psi^T_1 K \Gamma \psi_2$ and the iso-triplet meson operators of the form $\bar{\psi} \Gamma \psi$.

Strictly speaking, Eq.~\eqref{eq:diquark_corr} should be understood as a statement about the connected part, where the diquark two-point function is represented by the connected Wick contraction alone. In the superfluid phase, however, the iso-singlet scalar-diquark operator can acquire a nonzero expectation value at $j\ne 0$, and hence the full correlator can receive a contribution from disconnected diagrams. This subtlety is important in lattice simulations, but it does not alter the conclusion that the lightest state belongs to the iso-singlet scalar diquark sector.

\subsection{Chiral symmetry breaking and correlation functions for chiral partners}\label{sec:chiral-partner}
So far we have focused mainly on how to determine hadron masses.  We now consider the restoration of chiral symmetry at finite density by examining two-point correlation functions for chiral partners.

In our previous work, we studied the $\mu$-dependence of the chiral condensate (see Fig.~6 in Ref.~\cite{Iida:2024irv}) using the same lattice setup as summarized in the next section. 
As long as the system stays in the hadronic phase, the chiral condensate remains independent of $\mu$. 
Immediately after the transition to the superfluid phase, the system is in the BEC regime where the condensate decreases as $\langle \bar{q}q \rangle \propto 1/\mu^2$, consistent with the prediction of ChPT~\cite{Kogut:2000ek}. 
In the BCS regime, it becomes approximately constant as a function of $\mu$. 
Although this behavior suggests a tendency toward chiral symmetry restoration, the plateau takes a large nonzero value in our lattice simulation.
This is partly due to the use of Wilson fermions, for which an additive renormalization subtraction is required to extract the physical condensate.
Furthermore, in our simulation, we take a relatively large value of the quark mass, so that it is natural that the chiral symmetry is not exactly restored.

As an alternative diagnostic to investigate the chiral restoration, one can rely on two-point correlation functions for chiral partners. 
If chiral symmetry is restored, then the correlators for chiral partners must become degenerate at each spacetime separation. 
Below, we briefly summarize the theoretical basis of this statement~\cite{Hayano:2008vn}.

As reviewed in Sec.~\ref{sec:flavor-sym}, massless $N_f=2$ three-color QCD at $\mu=0$ has an $SU(2)_L \times SU(2)_R \times U(1)_B$ symmetry in the Lagrangian, but in the vacuum it only retains $SU(2)_V \times U(1)_B$.
Indeed, we can define the generators of vector and axial-vector $SU(2)$ rotations by recombining $SU(2)_L \times SU(2)_R$, but the vacuum, $\ket{0}$, is not invariant under the $SU(2)_A$ rotation if chiral symmetry is spontaneously broken. This vacuum property can be represented as
\beq
Q^a \ket{0} = 0, 
\qquad 
Q_5^a \ket{0} \neq 0,
\eeq
where
\beq
Q^a(t) = \int d^3 x \, V_0^a(t,\vec{x}), 
\qquad 
Q_5^a(t) = \int d^3 x \, A_0^a(t,\vec{x}) \label{eq:def-vac}
\eeq
are the generators associated with the triplet vector ($V_0^a = \bar{\psi}\gamma_0 \tau^a \psi$) and axial-vector ($A_0^a = \bar{\psi}\gamma_0 \gamma_5 \tau^a \psi$) currents. Here, the adjoint indices of $SU(2)$ run $a=1,2,3$, and $\tau^a$ acts on the flavor indices~\footnote{In this discussion we assume the Minkowski spacetime. 
We replace $\gamma_0 \rightarrow i\gamma_4$ if we consider the Euclidean correlators in lattice calculations.}.

We also denote the scalar and pseudoscalar operators as
\beq
S^{0}&=&\bar{\psi}\psi, \quad S^{a}=\bar{\psi}\tau^a \psi,\\
P^{0}&=&\bar{\psi}i \gamma_5 \psi, \quad P^{a}=\bar{\psi} i\gamma_5 \tau^a \psi.
\eeq
Then, the quark-bilinear operators follow the commutation relations:
\begin{align} 
[Q^a_5(t), V_\mu^b (t,\vec{x})] &= i\epsilon_{abc}A_\mu^c(t,\vec{x}), \quad [Q^a_5(t), A_\mu^b (t,\vec{x})]=i\epsilon_{abc} V_\mu^c(t,\vec{x}), \\
{}[Q^a_5(t), S^0 (t,\vec{x})] &= iP^a(t,\vec{x}), \quad \quad ~[Q^a_5(t), P^0 (t,\vec{x})]= -i S^a(t,\vec{x}), \\
{}[Q^a_5(t), S^b (t,\vec{x})] &= i\delta_{ab}P^0(t,\vec{x}), \quad [Q^a_5(t), P^b (t,\vec{x})]=-i \delta_{ab}S^0(t,\vec{x}).
\end{align}
If the chiral symmetry is unbroken, then the vacuum expectation values of all the following commutators vanish:
\beq
\langle [ Q_5^{a_n}, \cdots [ Q_5^{a_2},[Q_5^{a_1},\Phi] \cdots] \rangle_0 =0
\eeq
for an arbitrary operator $\Phi$.
Taking $\Phi= S^0(x) S^0(y), P^a (x) P^b(y)$ and so on, we conclude
\beq
\langle S^0(x) S^0(y) - P^a(x) P^a(y) \rangle_0 = 0,
\label{eq:S-P-relation}
\\
\langle V_\mu^a(x) V_\nu^a(y) - A_\mu^a(x) A_\nu^a(y) \rangle_0 = 0,
\label{eq:V-A-relation}
\eeq
if $Q_5^a \ket{0} = 0$ or, equivalently, if the chiral symmetry is restored.
These relations hold at each spacetime point $x$ and $y$. In other words, if the chiral symmetry is restored, the two-point correlation functions for chiral partners become degenerate.

For $N_f=2$ QC$_2$D, the above discussion also applies.
Indeed, the $SU(2)_V$ symmetry remains unbroken even in the superfluid phase. 
Therefore, the operator structures of $V_\mu^a$ and $A_\mu^a$ 
are unchanged from those in ordinary QCD.
Recall, furthermore, that we introduce the diquark source term~\eqref{eq:diquark-source-term} into the Lagrangian to control the $U(1)_B$ breaking.
One finds that $\mathcal{L}_j$ is invariant under an infinitesimal $SU(2)_A$ transformation.

What differs in QC$_2$D and in the superfluid phase in particular?
Even if the chiral symmetry is restored, one should note that the $U(1)_B$ symmetry is spontaneously broken in the superfluid phase. 
As a consequence, the spectrum created by the mesonic operators $S^0$ and $A_{\mu}^a$ contains not only mesonic modes but also baryon-type modes due to meson--baryon mixing as discussed in Sec.~\ref{sec:meson-baryon-mixing}. 
Therefore, neither the $S^0$--$S^0$ nor the $A_\mu^a$--$A_{\mu}^a$ correlator represents a purely mesonic excitation. 
Nevertheless, the relations~\eqref{eq:S-P-relation} and \eqref{eq:V-A-relation} can still hold as a statement about the full correlators for mixed excitations, provided that the chiral symmetry is restored.

We add a remark here. As in three-color QCD, the axial-vector correlator can receive a spin-zero pion contribution through the PCAC relation, in particular through the temporal component of $A_\mu^a$. 
When comparing with the vector current, therefore, it is necessary to remove the $J=0$ contribution 
~\footnote{In a similar way, in the superfluid phase where $U(1)_B$ is spontaneously broken, the iso-singlet vector current 
$J_\mu^B=\sum_f \bar{\psi}_f \gamma_\mu \psi_f$ 
receives a similar contribution as far as its temporal component is concerned.  Note, however, that this belongs to a different channel from those in Eq.~\eqref{eq:V-A-relation}. }.
The issue does not arise if one considers the spatial components ($\mu=i$) in Eq.~\eqref{eq:V-A-relation}. 
In this work, we thus focus on $V_i^a$ and $A_i^a$.

\section{Lattice setup}\label{sec:setup}

\subsection{Lattice action}
\label{subsec:action}
We briefly summarize the lattice setup for $N_f=2$ QC$_2$D used in a series of our studies~\cite{Iida:2019rah, Iida:2020emi,Iida:2022hyy, Ishiguro:2021yxr, Iida:2024irv}, where we use the Iwasaki gauge action~\cite{Iwasaki:1983iya} and Wilson fermions.
For the lattice fermion action, we introduce a diquark source term as
\beq
S_F= \bar{\psi}_1 \Delta(\mu)\psi_1 + \bar{\psi}_2 \Delta(\mu) \psi_2 - J \bar{\psi}_1 K \bar{\psi}_2^{T} + \bar{J} \psi_2^T K \psi_1,\label{eq:action}
\eeq
with $J=j \kappa$ and $\bar{J}= \bar{j} \kappa$ corresponding to the antidiquark and diquark source parameters, respectively. Here, the factor $\kappa$, a hopping parameter, comes from the rescaling of the Wilson fermion on the lattice. 
The index $f=1,2$ of $\psi_f$ denotes the flavor, and $\Delta(\mu)$ denotes the Wilson-Dirac operator including the term proportional to the quark number operator.  At $\mu>0$,
$\Delta(\mu)$ breaks the $\gamma_5$-hermiticity in three-color QCD, but in QC$_2$D, where the pseudo-reality of the fundamental representation of $SU(2)$, Eq.~\eqref{eq:pseudoreal}, holds, the Wilson-Dirac operator satisfies $[\det \Delta (\mu)]^* = \det \Delta (\mu)$. Then, there is no sign problem as long as we consider the $N_f=2$ case~\cite{Muroya:2000qp,Cohen:2003kd, Muroya:2003qs}.
For simplicity, we put $J=\bar{J}$ and take it to be real. 
The diquark source term serves as an external source that explicitly fixes the direction of $U(1)_B$ breaking.
Spontaneous symmetry breaking can then be diagnosed by extrapolating the $j \to 0$ limit and examining whether the expectation value of the associated operator remains nonzero.

To incorporate this action into the HMC framework, we rewrite Eq.~\eqref{eq:action} using an extended fermion matrix $\mathcal M$ as
\beq
S_F&=& (\bar{\psi}_1 ~~ -\psi_2^T C T^2) \left( 
\begin{array}{cc}
\Delta(\mu) & J \gamma_5 \\
-J \gamma_5 & \Delta(-\mu) 
\end{array}
\right)
\left( 
\begin{array}{c}
\psi_1  \\
C^{-1} T^2 \bar{\psi}_2^T 
\end{array}
\right)
 \equiv  \bar{\Psi}' {\mathcal M} \Psi'.  \label{eq:def-M}
\eeq
Thanks to the pseudo-reality of the fundamental fermions in the $SU(2)$ gauge theory, $\psi_1$ and the charge conjugation of $\bar{\psi}_2^T$ can be put in the same multiplet. 
For real $J(=\bar{J})$, the squared extended matrix becomes block diagonal.   
We thus obtain
\beq
\det[{\mathcal M}^\dag {\mathcal M}] = 
\det[ \Delta^\dag(\mu)\Delta(\mu) + J^2 ]~\det[  \Delta^\dag (-\mu) \Delta(-\mu) + J^2  ]. \label{eq:MdagM}
\eeq
From the point of view of actual numerical calculations, the $J$ insertion speeds up calculations since it lifts the eigenvalues of the matrix~\cite{Kogut:2001na, Kogut:2002cm,Skullerud:2003yc}.
To simulate two flavors, we take the square root of $\det[{\mathcal M}^\dag {\mathcal M}]$ corresponding to the fermion determinant of the four-flavor theory and utilize the Rational Hybrid Monte Carlo (RHMC) algorithm in our numerical simulations.

\subsection{Fermion propagators}\label{sec:fermi-propagator}
To construct hadron correlation functions for degenerate quark masses, we first define the fermion propagator from the fermion action~\eqref{eq:def-M}.
Following the notation of Ref.~\cite{Hands:2007uc}, we rewrite the action in the form
\beq
S_F&=& \bar{\Psi} \tilde{\mathcal{M}}\Psi 
\eeq
for later use.
Here, we change the basis of the fermion multiplet from $\Psi'$ in Eq.~\eqref{eq:def-M} to
\beq
\bar{\Psi} &\equiv& (\bar{\psi}_1, \psi_2^T, \bar{\psi}_2, \psi_1^T ),\quad  \Psi \equiv( \psi_1, \bar{\psi}_2^T, \psi_2,\bar{\psi}_1^T)^T 
\eeq
in such a way that $\tilde{\mathcal{M}}$ is block-diagonalized as
\beq
\tilde{\mathcal{M}}&=&
\begin{pmatrix}
A &  0 \\
0  & \bar{A}  \\
\end{pmatrix}, \qquad
A= \frac{1}{2}
\begin{pmatrix}
\Delta(\mu) &  -JK \\
JK & -\Delta (\mu)^T  \\
\end{pmatrix},
 \qquad
\bar{A}= \frac{1}{2}
\begin{pmatrix}
\Delta(\mu) &  JK \\
-JK & -\Delta (\mu)^T  \\
\end{pmatrix}.\nonumber\\
\eeq

Then, the propagator can be written as
\beq
\contraction[1ex]{}{\Psi}{(x)}{\bar{\Psi}}
\Psi(x)\bar{\Psi}(y)
= [\tilde{\mathcal{M}}^{-1}](x,y)=
\begin{pmatrix}
S_N & -S_A & 0 & 0 \\
\bar{S}_A & \bar{S}_N & 0 & 0 \\
0 & 0 & S_N & S_A \\
0 & 0 & -\bar{S}_A & \bar{S}_N
\end{pmatrix},
\eeq
where the color and spinor indices are suppressed.
Here, $S_N$ and $S_A$ denote the normal and anomalous propagators, respectively.
Explicitly, they are given by 
\beq \label{eq:props_wrt_cont}
S_{N}&=&\contraction[1ex]{}{\psi}{_f(x)}{\bar{\psi} }
\psi_f(x) \bar{\psi}_f(y)=  Q^{-1}(\mu) \Delta^{\dag}(\mu), 
\quad
\bar{S}_N= \contraction[1ex]{}{\psi}{_f^T(x)}{\bar{\psi} }
\psi_f^T(x) \bar{\psi}_f^T(y)= (K\gamma_5)  Q^{-1}(-\mu) \Delta^\dag(-\mu)  (K\gamma_5), \nonumber \\
S_A &=& \contraction[1ex]{}{\psi}{_2(x)}{\psi}
\psi_2(x) \psi_1^T(y)=J Q^{-1}(\mu) K,
\quad~~
\bar{S}_A = \contraction[1ex]{}{\bar{\psi}}{_2(x)}{\bar{\psi}}
\bar{\psi}_2^T(x) \bar{\psi}_1(y)=J  (K\gamma_5)  Q^{-1}(-\mu) \gamma_5 ,
\eeq
where $Q(\mu)=\Delta^\dag(\mu) \Delta(\mu) + J^2$. 
In the continuum theory, $S_A$ arises in the superfluid phase from particle--hole mixing.
Since it is proportional to the source parameter $j$, it vanishes in the $j\to0$ limit  at finite volume.
In practical lattice calculations, the correlation functions for the hadrons are first computed at finite $j$, from which the hadron masses are extracted for each value of $j$, and then the  $j\to0$ limit is taken to obtain the final results for the masses.
In this way, the contribution of the anomalous propagator in the superfluid phase can be incorporated.

\subsection{Correlation functions for hadrons}\label{sec:had-corr}
We now express the hadron correlation functions in terms of the fermion propagators.
To make the present paper self-contained, we write them explicitly here, following Ref.~\cite{Hands:2007uc}.

As for the iso-singlet ($I=0$) meson operators defined by $M^0 = (\bar{\psi}_1 \Gamma \psi_1 +  \bar{\psi}_2 \Gamma \psi_2)/\sqrt{2}$ in Table~\ref{table:hadron-ops}, their two-point correlation functions are given by
\beq
\langle M^0(\tau) M^{0 \dag}(0) \rangle_F = 
&-&2 \Tr[S_N(\tau,\vec{x}|\tau,\vec{x}) \Gamma]^{\rm disc.}
\Tr[S_N(0,\vec{y}|0,\vec{y})\bar{\Gamma}]^{\rm disc.} \nonumber
\\
&+& \Tr[S_N(\tau,\vec{x}|0,\vec{y}) \bar{\Gamma}
S_N(0,\vec{y}|\tau,\vec{x})\Gamma] \nonumber
\\
&+& \Tr[S_A(\tau,\vec{x}|0,\vec{y}) \bar{\Gamma}^T
\bar{S}_A(0,\vec{y}|\tau,\vec{x})\Gamma].\label{eq:2ptfunc_I0meson}
\eeq
Here, $\Tr [\cdot]$ denotes the trace over color and spinor indices together with the sums over the spatial coordinates $\vec{x}$ and $\vec{y}$ for zero-momentum projection.
The subscript $F$ labels the fermion contraction~\cite{Gattringer:2010zz}.
The first term on the right-hand side corresponds to the disconnected contribution. 
More precisely, it can be calculated by the correlation between one-point functions $O(\tau,\vec{x})$ and $O(0,\vec{y})$, where the vacuum expectation value of each one-point function is subtracted as~\footnote{Because of this subtraction, the two-point correlation function defined here does not have the constant term originated from the one-point function.} 
\beq
\Tr[O(\tau,\vec{x})]^{\rm disc.} \coloneqq  {\Tr}
\left[
O(\tau,\vec{x})
\right]
-
\left\langle
\frac{1}{N_\tau}
\sum_\tau
{\rm Tr}
[O(\tau,\vec{x})]
\right\rangle.
\eeq

The iso-triplet meson operators ($I=1$) consist of a neutral meson, $ (\bar{\psi}_1 \Gamma \psi_1 -  \bar{\psi}_2 \Gamma \psi_2)/\sqrt{2}$, and charged mesons, $M^1 = \bar{\psi}_1 \Gamma \psi_2$ and  $\bar{\psi}_2 \Gamma \psi_1 $.
Taking the fermion contractions, one finds that the two-point correlation functions for both types of mesons have the same expression:
\beq
\langle M^1(\tau) M^{1 \dag}(0) \rangle_F 
&=& 
\Tr[S_N(\tau,\vec{x}|0,\vec{y}) \bar{\Gamma} 
S_N(0,\vec{y}|\tau,\vec{x})\Gamma]
\nonumber
\\
&-& \Tr[S_A(\tau,\vec{x}|0,\vec{y}) \bar{\Gamma}^T
\bar{S}_A(0,\vec{y}|\tau,\vec{x})\Gamma].\label{eq:2ptfunc_pirho}
\eeq
For these iso-triplet mesons, disconnected diagrams do not contribute to the two-point functions. 
Note that the second term has the opposite sign to that of the third term in Eq.~(\ref{eq:2ptfunc_I0meson}) for $I=0$ mesons.

The iso-singlet diquark operators defined by $D^0 =( \psi^T_1 K \bar{\Gamma} \psi_2 -  \psi^T_2 K \bar{\Gamma} \psi_1)/\sqrt{2}$ lead to the two-point correlation functions:
\beq
\langle D^0(\tau) D^{0 \dag}(0) \rangle_F 
=
&-&2\Tr[\bar{S}_A(\tau,\vec{x}|\tau,\vec{x}) \Gamma K]^{\rm disc.}
\Tr[S_A(0,\vec{y}|0,\vec{y}) K \bar{\Gamma}]^{\rm disc.} \nonumber
\\
&-& \Tr[S_N(\tau,\vec{x}|0,\vec{y}) \Gamma K 
\bar{S}_N(0,\vec{y}|\tau,\vec{x})K\bar{\Gamma}] \nonumber
\\
&-& \Tr[S_N(\tau,\vec{x}|0,\vec{y}) K \Gamma^T
\bar{S}_N(0,\vec{y}|\tau,\vec{x})K\bar{\Gamma}].\label{eq:2ptfunc_I0diquark}
\eeq
The first term of the right-hand side, which corresponds to the disconnected diagram composed of the anomalous propagators, can take nonzero values in the superfluid phase.

In the iso-triplet diquark channel, only $J^P = 1^+$ axial-vector diquarks are allowed among the diquark states considered here.  Nevertheless, writing the expression in terms of a general $\Gamma$, the two-point function reads 
\beq
\langle D^1(\tau) D^{1 \dag}(0) \rangle_F &=& 
\Tr[S_N (\tau,\vec{x}|0,\vec{y})\Gamma K \bar{S}_N(0,\vec{y}|\tau,\vec{x})K \bar{\Gamma}]
\nonumber \\ &&
-\Tr[S_N (\tau,\vec{x}|0,\vec{y})K\Gamma^T \bar{S}_N(0,\vec{y}|\tau,\vec{x})K \bar{\Gamma}]. 
\label{eq:2ptfunc_I1diquark}
\eeq
The disconnected diagram does not contribute to the two-point functions.
For the connected parts, the relative sign differs from that of the connected part for $I = 0$ diquarks.

Finally, the two-point functions of the NG mode ($D^-$) associated with the spontaneous breaking of $U(1)_B$ in the superfluid phase and of the Higgs mode ($D^+$) orthogonal to $D^-$ are given as follows:
\beq
\langle D^{\pm}(\tau) D^{\pm \dag}(0) \rangle_F &=&  \mp \Tr[K \bar{S}_A(\tau,\vec{x}|\tau,\vec{x})]^{\rm disc.}  \Tr[K \bar{S}_A(0,\vec{y}|0,\vec{y})]^{\rm disc.}   \nonumber\\
&&  \mp \Tr[K S_A(\tau,\vec{x}|\tau,\vec{x})]^{\rm disc.} \Tr[K S_A(0,\vec{y}|0,\vec{y})]^{\rm disc.}   \nonumber\\
&&   - 2 \Tr[K S_A(\tau,\vec{x}|\tau,\vec{x})]^{\rm disc.} \Tr[K \bar{S}_A(0,\vec{y}|0,\vec{y})]^{\rm disc.}   \nonumber\\
&& \pm \Tr[\bar{S}_A(\tau,\vec{x}|0,\vec{y}) K \bar{S}_A(0,\vec{y}|\tau,\vec{x}) K] \nonumber\\
&& \pm \Tr[S_A(\tau,\vec{x}|0,\vec{y}) K S_A(0,\vec{y}|\tau,\vec{x}) K] \nonumber\\
&& - \Tr[S_N(\tau,\vec{x}|0,\vec{y}) K \bar{S}_N(0,\vec{y}|\tau,\vec{x}) K] \nonumber\\
&&- \Tr[\bar{S}_N(\tau,\vec{x}|0,\vec{y}) K S_N(0,\vec{y}|\tau,\vec{x}) K]. 
\label{eq:2ptfunc_higgs}
\eeq
Here, the first three lines represent the disconnected contributions.
In the hadronic phase, the anomalous propagator vanishes, so that these two modes are completely degenerate. In the superfluid phase, it is expected that the NG mode becomes lighter, while the Higgs mode becomes heavier.

\subsection{Simulation parameters and methods}
\label{sec:sim-params}
In this work, we use the same gauge configurations as those used in Ref.~\cite{Iida:2024irv}, where the phase structure and equation of state were investigated.
We take ($\beta, \kappa, N_s, N_\tau$) $=$ ($0.800,0.159,32,32$), which corresponds to  $T = 0.19 T_c = 40$ MeV with $a=0.17$ fm if we set the pseudo-critical temperature, $T_c$, at $\mu=0$ to $200$ MeV to fix the physical scale.
We impose periodic boundary conditions in the spatial directions and anti-periodic boundary conditions in the temporal direction. 
The simulations with ($\beta,\kappa$) $=$ ($0.800, 0.159$) give $m_{\pi}/m_\rho=0.813(1)$ and $am_{\pi}= 0.620(1)$ ($m_{\pi}\approx 738$ MeV) ~\cite{Murakami:2022lmq, Murakami:2023ejc}, which will be confirmed in this work~\footnote{In previous works, we denoted the pseudoscalar (PS) and vector (V) mesons in QC$_2$D as ``$m_{\rm PS}/m_V$" to avoid confusion with the ordinary pion and $\rho$ meson in three-color QCD. In this work, however, we simply refer to them as $\pi$ and $\rho$. }.

As for the quark chemical potential, $\mu$, we take $0 \le a\mu \le 0.75$.
According to our previous work on the phase diagram~\cite{Iida:2024irv}, the hadronic-superfluid phase transition occurs at $\mu_c/m_{\pi}=0.47$ or, equivalently, $a\mu_c=0.29$.
The guiding line for the BEC-BCS crossover is located at $\mu^{\rm BEC/BCS}/m_\pi = 0.73$, which is $a\mu^{\rm BEC/BCS}= 0.45$ in lattice units.
For the diquark source parameter $aj$, configurations are generated at $aj=0$ for $a\mu \le 0.25$, while $aj = 0.010, 0.015$, and $0.020$ are employed for $a\mu \ge 0.27$.
After measuring physical observables at each $aj$, we extrapolate them to the $j \rightarrow 0$ limit at each fixed value of $\mu$.

In the measurements of two-point functions, we use wall-type quark operators at the source and local quark operators at the sink for the connected parts.
Regarding the disconnected parts, we evaluate the traces of the relevant one-point operators using the random noise method with complex U(1) random numbers~\cite{Ejiri:2009hq}. We use one noise vector for each color and spinor index.

For each ($a\mu,aj$), we measure $100$--$700$ configurations, separated by $5$--$20$ HMC trajectories.
All statistical errors are estimated by the jackknife method.

\section{Results for hadron spectra}\label{sec:results}
In this section, we present the results for the two-point correlation functions, $C(\tau)$, for hadrons and the corresponding effective masses extracted from them.
The correlation-function data for all channels analyzed in this work are publicly available on Zenodo~\cite{Iida:2026dataset}.
We first describe the strategy for determining the effective masses.

For each set of parameters $(a\mu, aj)$, we extract the effective mass of each hadron from the two-point correlation function by assuming a fitting ansatz. We employ two types of fitting functions, namely, the cosh-type and exp-type functions, both of which contain two fitting parameters. Using two local data points in $\tau$ for $C(\tau)$, we determine these two parameters for each jackknife sample. Thus, what we do in practice corresponds to solving equations rather than performing a conventional fit. The resulting effective mass $m_{\rm eff.}(\tau)$ is then plotted as a function of $\tau$. We identify a plateau region, where we perform a constant fit over several data points to obtain $m_{\rm eff.}$ at each $(a\mu, aj)$. Finally, we extrapolate to $j \to 0$ to obtain the effective mass for each $\mu$.

The fitting functions are defined as follows.
At $\mu = 0$, due to the flavor symmetry discussed in Sec.~\ref{sec:flavor-sym} and the periodicity of lattices, 
the system exhibits a time-reversal symmetry under $\tau \leftrightarrow (N_\tau - \tau)$. Accordingly, the two-point correlation functions for all hadrons can be written as a linear combination of cosh-type functions. The effective mass of the lowest state in each channel can be extracted by fitting the correlator in the long-$\tau$ region to a single cosh function:
\beq
C(\tau) = c_0 \cosh [c_1 (\tau - N_\tau^{\mathrm{min}})].\label{eq:fit-fn-cosh}
\eeq
Here, $N_\tau^{\mathrm{min}}$ denotes the temporal position where the correlator takes its minimum.
For all hadronic operators at $\mu=0$, and also for meson correlation functions in the hadronic phase even at $\mu\neq0$, the time-reversal symmetry gives $N_\tau^{\mathrm{min}}=N_\tau/2$.
The fitted value of the parameter $c_1$ gives the resulting effective mass.

In contrast, once a nonzero quark chemical potential is introduced, the $\gamma_5$-hermiticity is lost, implying that the time-reversal symmetry is no longer guaranteed. In particular, 
the effective masses of the diquark and antidiquark states are linearly shifted by $\mu$ in the hadronic phase as shown in Eq.~\eqref{eq:mass-shift}. If the diquark contribution dominates the forward propagation while the antidiquark contribution does the backward propagation, then the $\tau \leftrightarrow (N_\tau - \tau)$ symmetry is explicitly broken (see, e.g., Fig.~\ref{fig:corr-D-Meson-1}). In such cases, we fit the correlator data only in the long-$\tau$ region using a single exp-type function:
\beq
C(\tau) = c_0 \exp (- c_1 \tau),\label{eq:fit-fn-exp}
\eeq
where the fitted value of the parameter $c_1$ gives the resulting effective mass.

In the superfluid phase, meson-baryon mixing occurs because of $U(1)_B$ symmetry breaking. Then, even for meson channels,  the time-reversal symmetry could be broken, so that the exp-type function may be more appropriate to extract the effective mass. 
On the other hand, as we will show, the two-point functions of diquarks (and antidiquarks) gradually exhibit a form consistent with a restored time-reversal symmetry at the highest densities considered here. 
This is partly because antidiquarks become sufficiently heavy to decouple and partly because, as discussed in Sec.~\ref{sec:meson-baryon-mixing}, our correlators do not resolve the meson-baryon mixing.
In that case, the lightest mode with a given quantum number may dominate both the forward and backward propagations, which
allows us to determine the effective mass from a cosh-type fitting function whenever it is supported by the correlator data.
In addition, in several heavy-hadron channels, the data around $\tau \sim N_\tau/2$ can be too noisy to allow a reliable cosh-type fit. In these cases, we adopt the exp-type fit.

For the $j \to 0$ extrapolation of the effective mass at each value of $\mu$ in the superfluid phase, we mainly use a linear function of $aj$, 
\beq
f(aj)= c_0+ c_1(aj),\label{eq:j-linear}
\eeq
and use three data points at $aj=0.010, 0.015$, and $0.020$ to fix $c_0$ and $c_1$.
For the NG mode, we also examine a square-root ansatz motivated by ChPT~\cite{Kogut:2000ek}~\footnote{In Ref.~\cite{Kogut:2000ek}, it is also predicted that the pion mode in the superfluid phase follows a similar functional form with $c_0\ne 0$.  In our data, however, the linear fit ansatz provides a better chi-squared. This may be due to the relatively heavy quark mass used in the present study, which makes it difficult to resolve the detailed $j$-dependence in channels with a finite mass such as the pion mode.}, 
\beq
g(aj)=c_0 + c_1 \sqrt{aj}.\label{eq:j-sqrt}
\eeq
It is natural to expect from the $j\to 0$ extrapolation that $c_0 \approx 0$ for the NG mode.

\subsection{Pion and rho meson}
We start with $\pi$ and $\rho$ mesons, which correspond to hadrons in isovector ($I = 1$), $J^P = 0^-$ and $J^P = 1^-$ channels. 
We calculate Eq.~\eqref{eq:2ptfunc_pirho} with $\Gamma = \gamma_5$ and $\gamma_i$ for the pion and the $\rho$ meson, respectively. Note that there is no contribution from disconnected diagrams.

\begin{figure}[htbp]
\centering
\includegraphics[width=.45\textwidth]{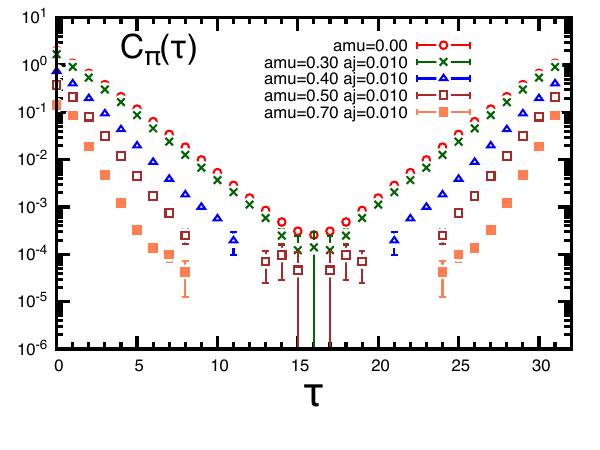}
\qquad
\includegraphics[width=.45\textwidth]{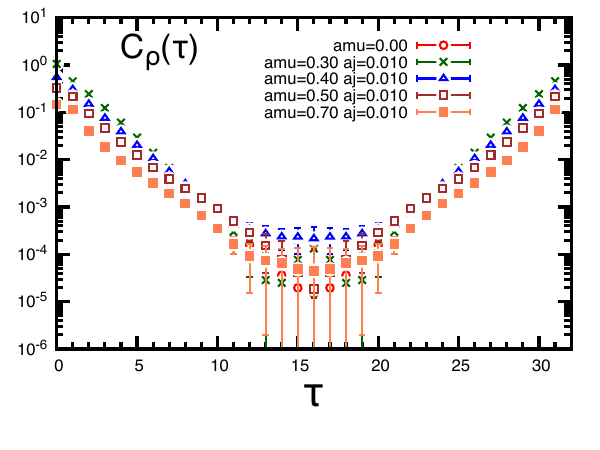}
\caption{The chemical potential dependence of the correlation function for $\pi$ (left) and $\rho$ (right) mesons. For the $\rho$ meson, the data at $a\mu=0.00$ and $0.30$ almost overlap. \label{fig:corr-Meson-4-6}}
\end{figure}
Figure~\ref{fig:corr-Meson-4-6} exhibits the results for the correlation functions for several values of $a\mu$ and $aj$. 
The left panel shows the pion channel.  
At $a\mu = 0.00$ (red-circle symbols), the pion is also the lightest hadron in QC$_2$D according to the QCD inequalities as discussed in Sec.~\ref{sec:qcd-inequality}. Indeed, we obtain clean data, which persist in the hadronic phase. 
With increasing $\mu$ in the superfluid phase, the slope of $\log(C_\pi(\tau))$ becomes steeper, indicating that the pion mass increases. For $a\mu \geq 0.40$ and $aj=0.010$, the values of $C_\pi(\tau)$ around $\tau \approx N_\tau/2$ are nearly zero, and the central values even become negative due to fluctuations; these data are not shown in the figure. In ordinary QCD, it is well known that for heavy hadrons the correlators become noisy at large $\tau$ when they couple to lighter multi-hadron states. We therefore suspect that a similar phenomenon may occur; for instance, the $P$-wave pion can decay into an axial-vector diquark and scalar diquark owing to $U(1)_B$ symmetry breaking. Indeed, the pion becomes much heavier in the superfluid phase, as we shall see below.

The right panel of Fig.~\ref{fig:corr-Meson-4-6} depicts the results for the $\rho$-meson channel.  
At $a\mu = 0.00$ (red-circle symbols), this meson also exhibits clean signals, since our quark mass is so heavy, $m_\pi/m_{\rho} =0.813(1)$, that the $\rho$-meson cannot decay into two pions.  This situation persists in the hadronic phase.
For $aj=0.010$, however, the correlation functions around $\mu_c$ ($a\mu=0.30, 0.40$) have gentler slopes in the short $\tau$ region compared to the one at $a\mu = 0.00$, and the value of $C_\rho(\tau)$ around $\tau = N_\tau/2$ is larger than that at $a\mu = 0.00$. 
This indicates that the $\rho$-meson becomes lighter in the finite-density region. 
As $a\mu$ increases further, the slope becomes even gentler, while around $\tau = N_\tau/2$, $C_\rho(\tau)$ suffers large fluctuations, as in the pion case. This suggests that the $\rho$-meson is not the lightest stable hadron in the superfluid phase.

Since these correlation functions are symmetric about $\tau = N_\tau/2$, we adopt the cosh-type fit ansatz~\eqref{eq:fit-fn-cosh} to extract the effective masses over the whole range of $a\mu$.
\begin{figure}[htbp]
\centering
\includegraphics[width=.6\textwidth]{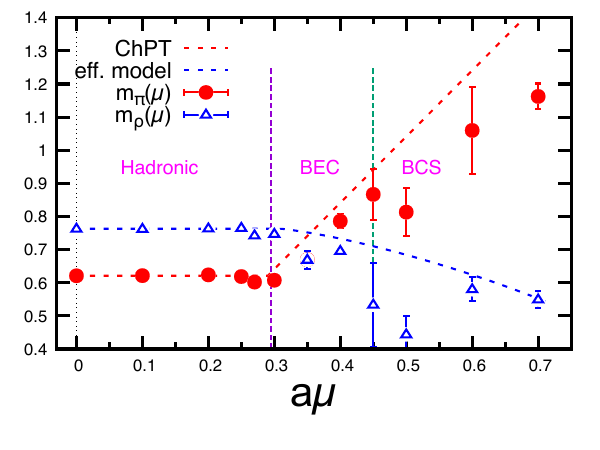}
\caption{The chemical potential dependence of the effective masses for the $\pi$ and $\rho$ mesons at $j \rightarrow 0$. At $a\mu=0.35$, the two masses are nearly degenerate and therefore overlap in the figure. 
The dashed line and curve show the analytical prediction for $m_\pi$ based on the ChPT~\cite{Kogut:2000ek} and the prediction for $m_\rho$ based on a linear sigma model that incorporates spin-1 excitations~\cite{Suenaga:2023xwa}, respectively.
The purple dashed line denotes the critical value, $a\mu_c=0.29$, which is the hadronic-superfluid
phase transition point, while the green dashed line indicates that the BEC-BCS crossover occurs around this value of $a\mu^{\rm BEC/BCS}=0.45$.
\label{fig:mass-Meson-4-6}}
\end{figure}
The $\mu$-dependence of the $\pi$- and $\rho$-meson masses after taking the $j\rightarrow 0$ limit is summarized in Fig.~\ref{fig:mass-Meson-4-6}.
The $\pi$-meson becomes heavier in the superfluid phase, while the $\rho$-meson in the superfluid phase becomes slightly lighter than that in the hadronic phase, as also pointed out in earlier studies~\cite{Muroya:2002jj, Muroya:2002ry, Hands:2007uc}. The ordering of these two masses is reversed between the hadronic and superfluid phases; more precisely, such an inversion occurs at around $a\mu = 0.35$.
We also find that the $\rho$-meson does not become massless even in the superfluid phase, while
the pion mass exceeds unity in lattice units at sufficiently high density.  The $m_{\pi}$ values at $a\mu = 0.60$ and $0.70$ may not be very reliable; nevertheless, we include them in Fig.~\ref{fig:mass-Meson-4-6} for reference.

In Fig.~\ref{fig:mass-Meson-4-6}, we also plot the analytical prediction based on ChPT for $m_\pi$, $m_{\pi}^{\rm ChPT}(\mu)=m_{\pi} (\mu=0) +2(\mu-\mu_c) \theta (\mu - \mu_c)$~\cite{Kogut:2000ek}, and the prediction for $m_\rho$ based on a linear sigma model that incorporates spin-1 excitations~\cite{Suenaga:2023xwa} as dashed line (red) and curve (blue), respectively~\footnote{In Fig.~\ref{fig:mass-Meson-4-6}, the blue dashed curve is an extract from Ref.~\cite{Suenaga:2023xwa}, where $g_\Phi=10$, $C=10$ and various physical inputs given by Eqs.~(31) and~(32) in Ref.~\cite{Suenaga:2023xwa} are adopted.}.
In our previous works~\cite{Iida:2019rah, Iida:2024irv}, lattice results for the diquark condensate and the equation of state in the BEC phase have shown a very good agreement with the ChPT predictions, whereas $m_\pi$ appears to deviate slightly from the ChPT behavior as compared to these observables. 
For $m_\rho$, it is not clear from $N_c=3$ QCD sum rules whether the $\rho$-meson becomes lighter or not~\cite{Hatsuda:1991ez, Hatsuda:1995dy, Hatsuda:1996az, Klingl:1997kf, Rapp:1999ej, Zschocke:2002mn,Steinmueller:2006id}, while in the two-color case,
according to an analysis by Lenaghan et al.~\cite{Lenaghan:2001sd} that partially incorporates spin-1 excitations, a downward-convex curve was predicted as shown in Fig.~2 of Ref.~\cite{Hands:2007uc}. 
In contrast, more recent predictions based on Refs.~\cite{Harada:2010vy,Suenaga:2023xwa} yield an upward-convex curve, as is consistent with that shown in Fig.~\ref{fig:mass-Meson-4-6}. Our lattice data suggest that $m_{\rho}$ decreases as $\mu$ increases, but determining its detailed behavior remains a task for future work.

\subsection{Higgs and NG modes}
\begin{figure}[htbp]
\centering
\includegraphics[width=.45\textwidth]{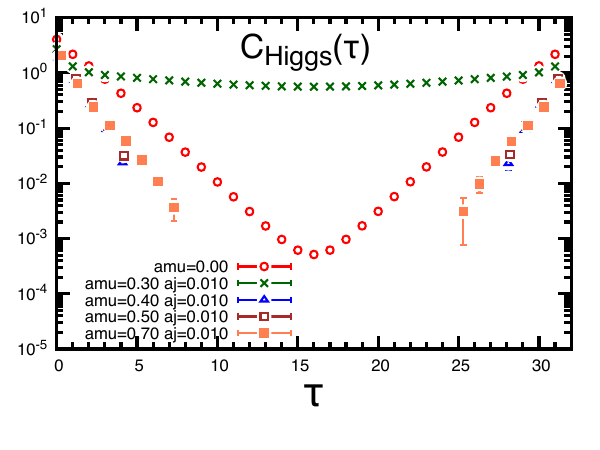}
\qquad
\includegraphics[width=.45\textwidth]{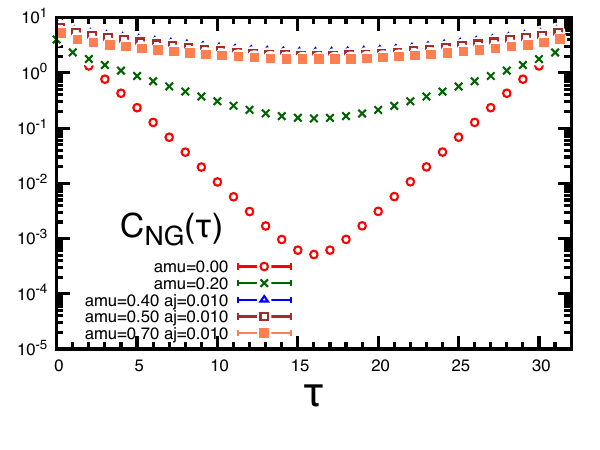}
\caption{The chemical potential dependence of the total correlation function for the Higgs mode (left) and NG mode (right). \label{fig:corr-Higgs}}
\end{figure}

Let us move on to the Higgs ($D^+$) and NG ($D^-$) modes in Table~\ref{table:hadron-ops}. 
The results for two-point correlation functions given by Eq.~\eqref{eq:2ptfunc_higgs} are shown in Fig.~\ref{fig:corr-Higgs}.
In the hadronic phase, the correlation functions for the Higgs (left panel) and NG (right panel) modes are completely identical. Although these operators are baryonic, no violation of time-reversal symmetry is observed. This is because the Higgs and NG operators are defined as linear combinations of diquark and antidiquark fields, for which the $\mu$-induced time-reversal asymmetry arises in the opposite direction and hence
cancels each other in the total correlator.
Once the system enters the superfluid phase, the NG mode  continues to exhibit a clean signal, whereas the Higgs mode becomes noisy as $\mu$  increases. At the same time, the slope of the Higgs-mode correlator becomes much steeper, which indicates that this mode becomes heavier in the high-density regime.

\begin{figure}[htbp]
\centering
\includegraphics[width=.45\textwidth]{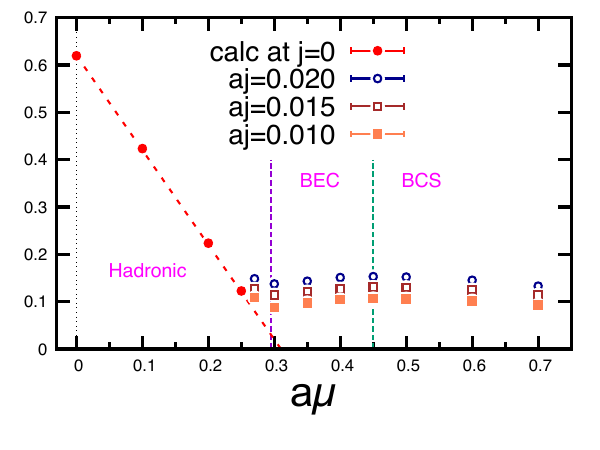}
\caption{The $\mu$- and $j$-dependences of the effective masses for the NG mode. \label{fig:NG-mass-j-deps}}
\end{figure}
Figure~\ref{fig:NG-mass-j-deps} depicts the effective masses of the NG mode for each ($a\mu, aj$). Here, we would like to focus on the results at finite $aj$, which exhibit a local minimum around $a\mu_c=0.29$ corresponding to $\mu/m_\pi \approx 0.5$. This behavior is consistent with the ChPT prediction (Eq.\ (101) and Fig.\ 3 of Ref.~\cite{Kogut:1999iv}). The fact that the mass of NG mode remains at $j \ne 0$ nonzero even at high density is also in agreement with this prediction~\footnote{Similar features have also been reported in analyses using staggered fermions~\cite{Wilhelm:2019fvp}.}.

\begin{figure}[htbp]
\centering
\includegraphics[width=.22\textwidth]{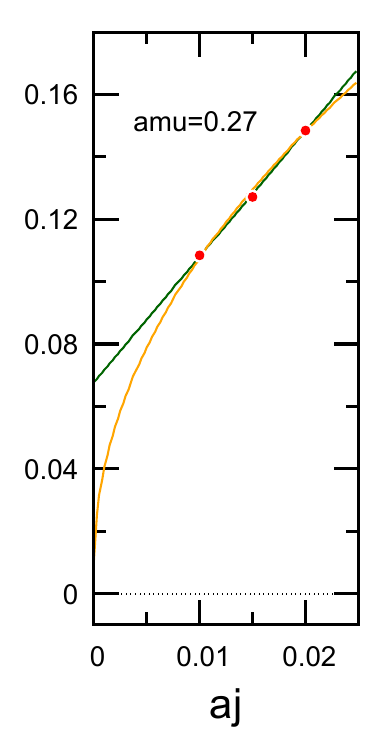}
\includegraphics[width=.22\textwidth]{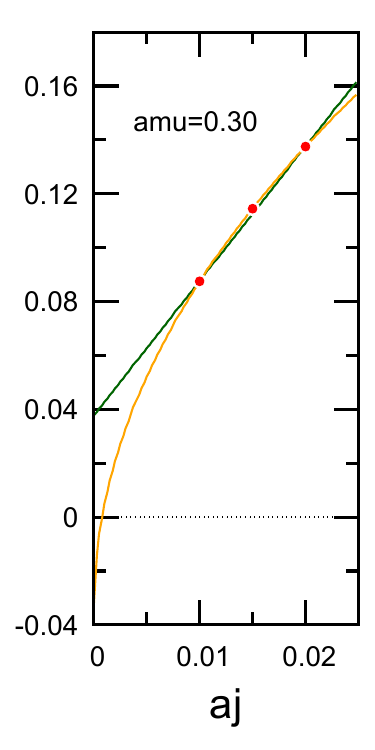}
\includegraphics[width=.22\textwidth]{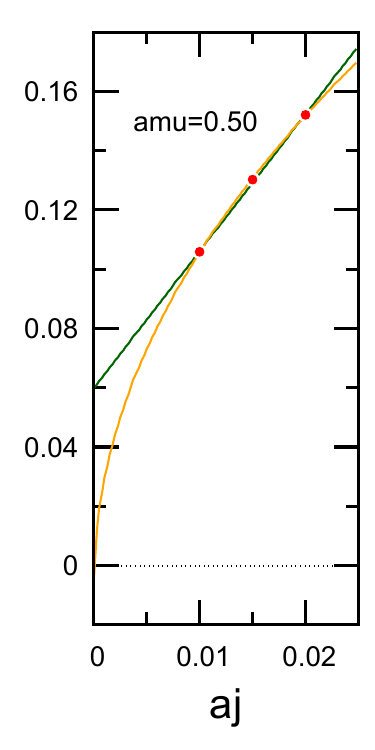}
\includegraphics[width=.22\textwidth]{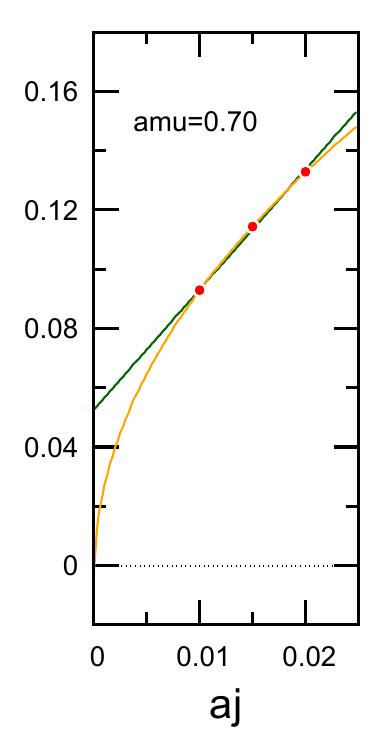}
\caption{Comparison between the linear- and sqrt-extrapolations in $(aj)$ for the effective mass of the NG mode. \label{fig:j0-extrpl-NG}}
\end{figure}

\begin{figure}[htbp]
\centering
\includegraphics[width=.45\textwidth]{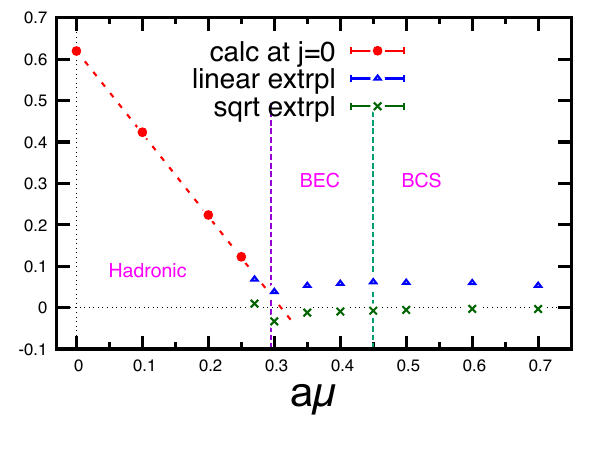}
\includegraphics[width=.45\textwidth]{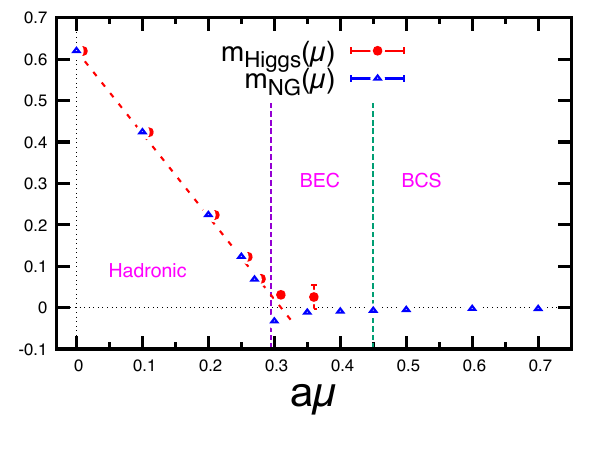}
\caption{(Left) Effective mass of the NG mode in the $j \to 0$ limit.
Red-circle symbols show direct data at $j = 0$, while blue-triangle and green-cross represent the results obtained by the linear- and sqrt-extrapolations, respectively. The red dotted line denotes $m(\mu)=m(\mu=0)-2\mu$.
(Right) The chemical potential dependence of the effective masses for the Higgs and NG modes in the $j \to 0$ limit. For $a\mu \ge 0.40$, the Higgs mode becomes too noisy to extract reliable mass data.}\label{fig:mass-Higgs}
\end{figure}

Let us now carefully examine the $j \to 0$ extrapolation. A comparison between the extrapolation using a linear function, Eq.~\eqref{eq:j-linear}, and that using a sqrt-type function, Eq.~\eqref{eq:j-sqrt}, motivated by the ChPT is shown in Fig.~\ref{fig:j0-extrpl-NG}. 
The fitting quality was measured by $\chi^2/{\rm dof}$, which amounts to 1.49, 3.61, 8.09, and 15.9 for the linear extrapolation and to 8.23, 0.03, 2.50, and 0.39 for the sqrt-type extrapolation at $a\mu = 0.27$, $0.30$, $0.50$, and $0.70$, respectively.
The results for these two extrapolations are summarized in the left panel of Fig.~\ref{fig:mass-Higgs}. 
At $a\mu = 0.27$, the system is close to but not yet in the superfluid phase. Not only does the linear extrapolation indicate a better fitting quality, but the extrapolated data (blue-triangle symbol) is consistent with the predicted mass-shift in the hadronic phase, Eq.~\eqref{eq:mass-shift}. 
For $a\mu \geq 0.30$, on the other hand, the sqrt-type extrapolation provides a better fitting. 
While the result at $a\mu = 0.30$ gives a slightly negative mass, for larger $a\mu$ the mass becomes nearly massless, which is consistent with the behavior expected of an NG boson.
We, therefore, adopt the linear extrapolation for $a\mu = 0.27$ and the sqrt-type extrapolation for $a\mu \geq 0.30$ to obtain our final results for the effective masses.
Note that such large systematic uncertainties associated with the extrapolation in $(aj)$ were also discussed in the previous study~\cite{Hands:2007uc}.

The resulting effective masses for the Higgs and NG modes are summarized in Fig.~\ref{fig:mass-Higgs}.
In the hadronic phase, the masses remain degenerate, scale as $m(\mu)=m(\mu=0)-2\mu$, which corresponds to the theoretical prediction, Eq.~\eqref{eq:mass-shift} with $n_O=+2$, and approach nearly zero around the onset of the superfluid transition. 
Beyond this point, the Higgs-mode correlator becomes too noisy to extract a reliable signal, while the NG mode remains massless.

\subsection{Iso-singlet scalar meson and diquark}
\begin{figure}[htbp]
\centering
\includegraphics[width=.45\textwidth]{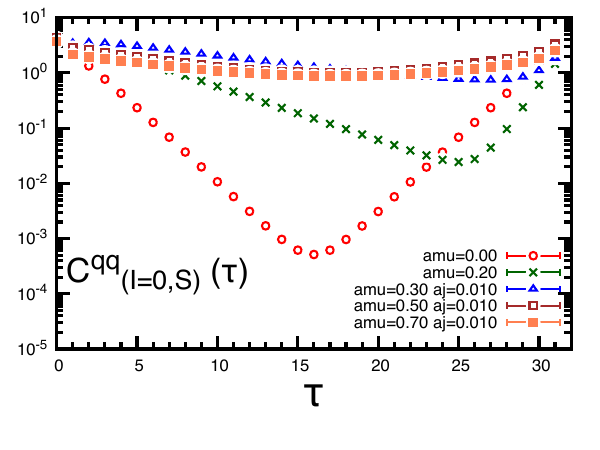}
\qquad
\includegraphics[width=.45\textwidth]{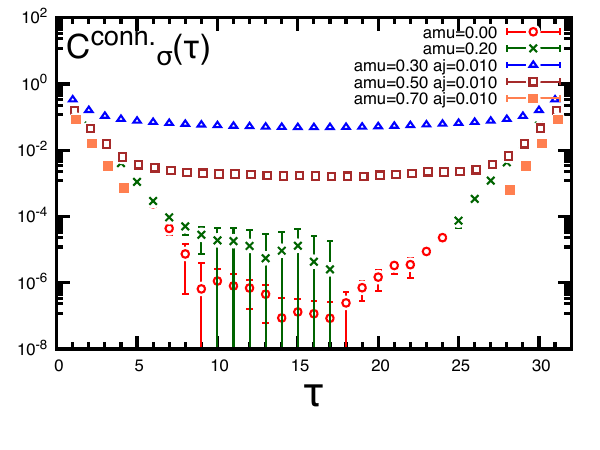}
\caption{The chemical potential dependence of the correlation function in the scalar diquark/antidiquark channel (left) and the $\sigma$-meson channel (right). Here, the $\sigma$-meson correlation function shows only the connected part of the two-point function, while the diquark correlation function includes the disconnected contribution. \label{fig:corr-D-Meson-1}}
\end{figure}

Next, we consider the diquark and meson channels with $I=0$, $J^{P}=0^{+}$, namely, the scalar diquark and $\sigma$-meson channels.
We first present the correlation functions for the diquark given by Eq.~\eqref{eq:2ptfunc_I0diquark} with $\Gamma = 1$ in the left panel of Fig.~\ref{fig:corr-D-Meson-1}. Here we interpret the forwardly propagating mode in Euclidean time $\tau$ as the diquark.
Accordingly, the backwardly propagating mode from $N_{\tau}=32$ can be regarded as the antidiquark.
In Fig.~\ref{fig:corr-D-Meson-1}, the red-circle data points correspond to the diquark correlator at $\mu=0$, which is symmetric under the time-reversal transformation $\tau \leftrightarrow (N_{\tau}-\tau)$. Once $\mu$ becomes nonzero, however, this symmetry is clearly violated~\cite{Muroya:2002ry, Murakami:2022lmq, Murakami:2023ejc}, as can be seen for example in the green-cross data obtained at $a\mu=0.20$. 
Once the system enters the superfluid phase ($a\mu \gtrsim 0.30$), the degree of time-reversal asymmetry becomes smaller as $\mu$ is increased further and eventually the diquark correlator approximately recovers a symmetric shape.
This behavior can be understood as a consequence of the meson-baryon mixing, here including baryon-antibaryon mixing, in the superfluid phase of QC$_2$D as explained in Sec.~\ref{sec:meson-baryon-mixing}.

Let us now turn to the right panel of Fig.~\ref{fig:corr-D-Meson-1}, which shows the connected correlator in the meson channel with the same quantum numbers as the scalar diquark, namely, the $\sigma$-meson.
In the hadronic phase, the meson correlator is expected to be time-reversal symmetric from the discussion in Sec.~\ref{sec:mass-shift}.
At $a\mu=0.20$ (green-cross symbols), however, the size of the statistical uncertainty is not symmetric around $\tau=N_\tau/2$.
The poor signal in this region may be related to two-particle modes that can couple to the $\sigma$ channel, such as the scalar diquark--antidiquark channel.
As will be seen later, at $a\mu=0.20$ the scalar diquark becomes very light, while the antidiquark mass also appears slightly smaller than the mass-shift prediction.
The $\sigma$-meson mass, within its large uncertainty, seems to lie close to or above the corresponding diquark--antidiquark threshold, which may make the connected $\sigma$ correlator noisy and difficult to interpret.

To extract the effective masses of the scalar diquark and antidiquark at small $\mu$, we assume an exp-type ansatz~\eqref{eq:fit-fn-exp} for the correlation since the correlator exhibits time-reversal asymmetry. Thus, we define the effective mass at each $\tau$ as
\beq
m_{\mathrm{eff.}} (\tau) =  \ln \frac{C(\tau)}{C(\tau +1)}.
\eeq
Also, in the regime $a\mu \gtrsim 0.30$, where the time-reversal asymmetry seems to be negligible, we adopt the cosh-type ansatz. 

\begin{figure}[htbp]
\centering
\includegraphics[width=.3\textwidth]{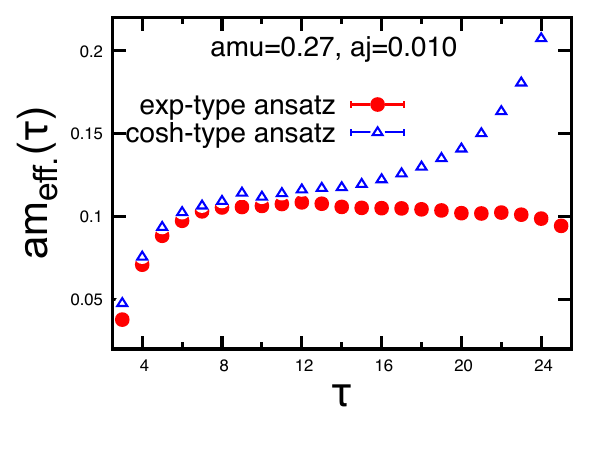}
\includegraphics[width=.3\textwidth]{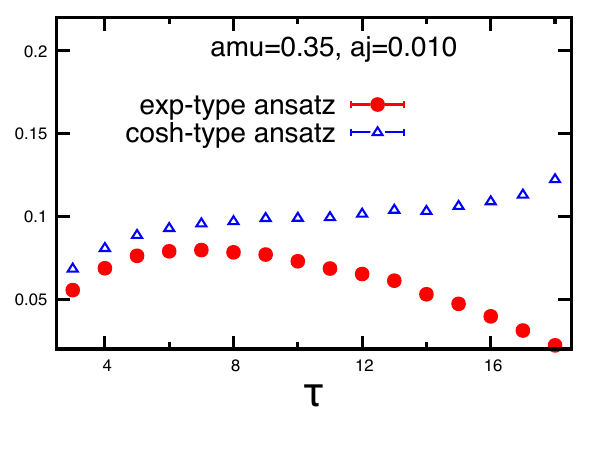}
\includegraphics[width=.3\textwidth]{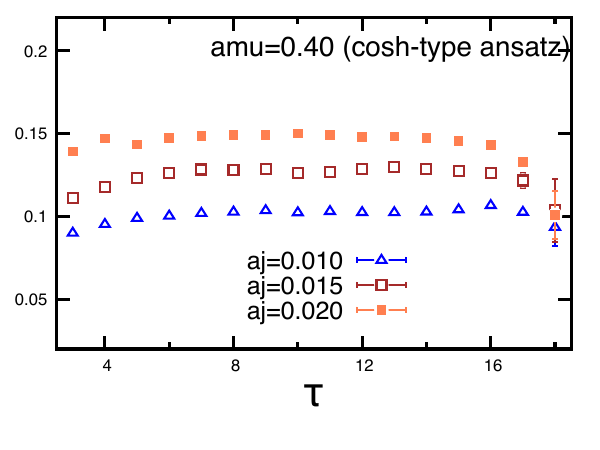}
\caption{The effective mass of the scalar diquark as a function of $\tau$. In the left ($a\mu=0.27, aj=0.010$) and middle ($a\mu=0.35,aj=0.010$) panels, we show the fitting-ansatz dependence.  The right panel shows the $j$-dependence at $a\mu=0.40$ using the cosh-type ansatz to obtain the effective masses.  } \label{fig:eff-mass-tau-Diquark-iqn1}
\end{figure}
In the left and middle panels of Fig.~\ref{fig:eff-mass-tau-Diquark-iqn1}, we compare the effective masses obtained using the two fitting ansatzes, exp-type and cosh-type, at $a\mu=0.27$ and $a\mu=0.35$.
At $a\mu=0.27$, the effective mass obtained with the exp-type ansatz exhibits a plateau over a wider $\tau$ range. The result based on the cosh-type ansatz shows a similar behavior to the exp-type one up to $\tau \lesssim 10$, while it starts to increase at even larger $\tau$.
Indeed, the correlator still exhibits a pronounced time-reversal asymmetry, and also the minimum of $C^{qq}_{(I=0,S)}(\tau)$ is located around $N^{\mathrm{min }}_\tau \simeq 27$ rather than at $\tau=N_\tau/2$.
Even the cosh-type ansatz explicitly given by $\cosh(c_1 (\tau - N^{\mathrm{min }}_\tau))$ would not match the data.
We conclude, therefore, that the exp-type ansatz provides a more reliable description of the data in this $\mu$-region. 

At $a\mu=0.35$, as can be seen from the middle panel of Fig.~\ref{fig:eff-mass-tau-Diquark-iqn1}, the situation is less clear. With the exp-type ansatz, one may still regard the region $6 \le \tau \le 8$ as an approximate plateau, but it forms a bump rather than a flat region. With the cosh-type ansatz, a more extended plateau-like behavior is observed around $9 \le \tau \le 11$. The main issue is that the plateau values extracted from these two ansatzes are significantly different. Therefore, we compute the effective masses using both ansatzes at three values of $j$, namely, $aj=0.010$, $0.015$, and $0.020$, and perform the extrapolation to the $j\to 0$ limit. Judging from the fit quality evaluated from, e.g., $\chi^2/{\rm dof}$ in the plateau fits and in the $j\to 0$ extrapolations, we find that the cosh-type ansatz is more appropriate at $a\mu=0.35$.
As $\mu$ is increased further, the plateau region obtained with the cosh-type ansatz becomes broader. As an example, we show, in the right panel of Fig.~\ref{fig:eff-mass-tau-Diquark-iqn1}, the results for the effective masses obtained at $a\mu=0.40$ and $aj=0.010, 0.015, 0.020$ using the cosh-type ansatz.
Based on the above discussions, we finally extract the effective masses at each ($a\mu, aj$) by using the exp-type ansatz for $a\mu \le 0.30$ and the cosh-type ansatz for $a\mu \geq 0.35$.

\begin{figure}[htbp]
\centering
\includegraphics[width=.45\textwidth]{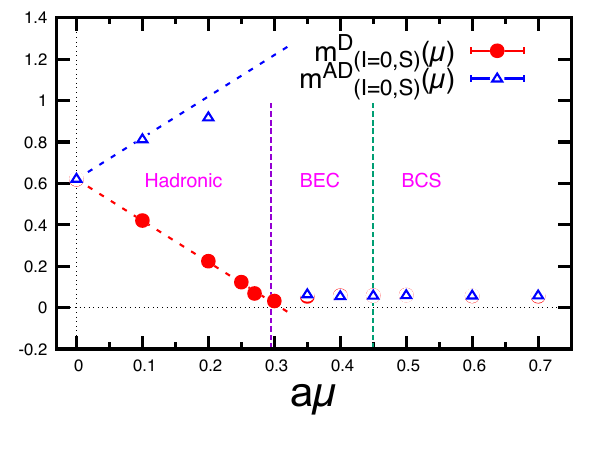}
\caption{The chemical potential dependence of the effective masses for the scalar diquark and antidiquark.
The dotted straight lines denote $m(\mu) = m(\mu=0) \pm 2\mu$, which are the theoretical predictions in Eq.~\eqref{eq:mass-shift}.}\label{fig:result-j0-mass-D-and-AD-iqn1}
\end{figure}
In Fig.~\ref{fig:result-j0-mass-D-and-AD-iqn1}, we display the diquark and antidiquark masses obtained by the linear $j\to 0$ extrapolation in Eq.~\eqref{eq:j-linear}.
In the hadronic phase, we also plot the theoretical mass-shift predictions, Eq.~\eqref{eq:mass-shift},
as dotted lines for comparison. As expected from the prediction, the diquark mass decreases linearly with increasing $\mu$ in the hadronic phase and becomes very small in the superfluid phase. 
The antidiquark mass in the hadronic phase also follows the mass-shift prediction, although a slight deviation is observed around $a\mu = 0.20$. 
As $\mu$ approaches the phase transition point $\mu_c$, the time-reversal asymmetry of $C^{qq}_{(I=0, S)}(\tau)$ becomes increasingly pronounced, which makes it difficult to identify a stable plateau for the effective mass of the antidiquark around $a\mu = 0.25$. Consequently, it is difficult to obtain a reliable value of the antidiquark mass just below $\mu_c$.
After the superfluid phase transition, the time-reversal asymmetry of the correlator gradually becomes smaller and eventually the effective mass of the antidiquark can again be extracted. 
In this regime, the antidiquark appears to become nearly degenerate with the diquark. This behavior can be interpreted as a consequence of the meson-baryon mixing, which allows mixing to occur among diquark, antidiquark, and mesonic excitations carrying the same quantum numbers. To clarify this point, one would need to examine the lowest-lying state in the mixed operator basis. 
A fully quantitative treatment of this mixing would require measuring the off-diagonal correlators between the diquark and meson operators and performing a variational analysis based on the generalized eigenvalue problem (GEVP). We leave such an extended analysis for future work.

\begin{figure}[htbp]
\centering
\includegraphics[width=.45\textwidth]{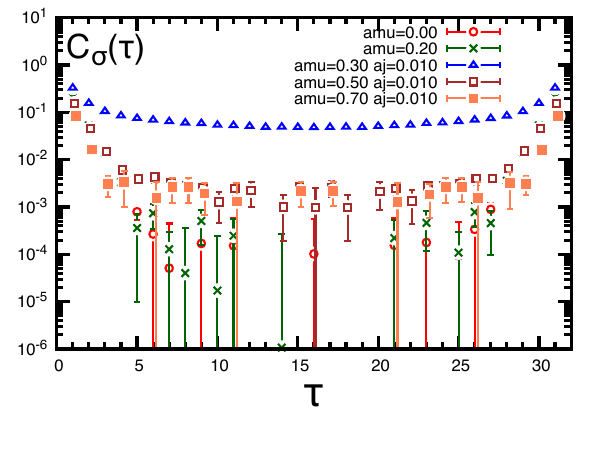}
\caption{The chemical potential dependence of the correlation function for the $\sigma$-meson, which includes the contributions from the disconnected part.  }\label{fig:total-corr-Meson-1}
\end{figure}
Finally, we study the meson channel with the same quantum numbers, namely, the $\sigma$-meson channel. The total correlator that has the disconnected contributions included is shown in Fig.~\ref{fig:total-corr-Meson-1}. By comparison with the diquark correlator in the left panel of Fig.~\ref{fig:corr-D-Meson-1}, one finds that the $\sigma$-meson correlator is much noisier over almost the entire $\mu$ range.
This originates from the structure of the fermion contractions. In the diquark channel, the disconnected contribution is written in terms of anomalous propagators, $S_A$, and is suppressed by powers of the small parameter $j$ (see Eq.~\eqref{eq:2ptfunc_I0diquark}). In contrast, in the meson channel, the disconnected contribution is expressed through the normal propagator, $S_N$, (see Eq.~\eqref{eq:2ptfunc_I0meson}), which can lead to much larger statistical fluctuations. The fact that the $\sigma$-channel is very noisy even at $\mu=0$ is consistent with the well-known difficulty in determining the $\sigma$ spectrum even in the standard vacuum of three-color QCD~\cite{Kunihiro:2003yj, Wakayama:2014gpa, Date:2026ngr}.

As $\mu$ increases, the $\sigma$-meson correlator becomes less noisy and the signal improves near the onset of the superfluid transition ($a\mu \approx 0.30$), indicating that the $\sigma$-meson becomes lighter in this region. For larger $\mu$ in the BCS phase, however, the slope of the correlator in the logarithmic plot becomes steeper, suggesting that the $\sigma$-meson becomes heavier again. In addition, particularly in the long-$\tau$ region, the correlator exhibits large fluctuations.

\begin{figure}[htbp]
\centering
\includegraphics[width=.45\textwidth]{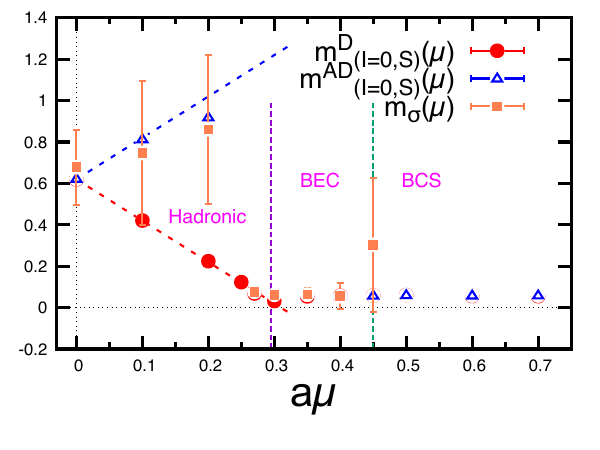}
\caption{The chemical potential dependence of the $\sigma$-meson mass (orange-squared symbols). For comparison, we also show the results for the scalar diquark and antidiquark masses presented in Fig.~\ref{fig:result-j0-mass-D-and-AD-iqn1}. \label{fig:eff-mass-Meson-iqn1}}
\end{figure}
To extract the $\sigma$-meson mass, we note that the total correlator is still symmetric with respect to $\tau \leftrightarrow (N_\tau-\tau)$. We therefore employ a cosh-type ansatz. After performing the extrapolation to the $j\to 0$ limit, the $\mu$-dependence of the resulting $\sigma$-meson mass is shown as orange-squared symbols in Fig.~\ref{fig:eff-mass-Meson-iqn1}. In the hadronic phase, the statistical errors of the mass data are very large, within which no significant $\mu$ dependence of the mass can be seen. In the superfluid phase, the $\sigma$-meson appears to become nearly degenerate with the diquark state at least in the relatively low-$\mu$ (BEC) region. As $\mu$ increases further, however, the statistical errors of the mass data grow again, making it difficult to obtain a reliable mass value for $a\mu \gtrsim 0.50$. As discussed above, a more systematic study of this channel would benefit from a variational analysis, which we leave for future work.

\subsection{Iso-triplet axial-vector meson and diquark}
Next, we consider the iso-triplet axial-vector channel, where the diquark correlator exhibits a relatively clean signal.  
\begin{figure}[htbp]
\centering
\includegraphics[width=.45\textwidth]{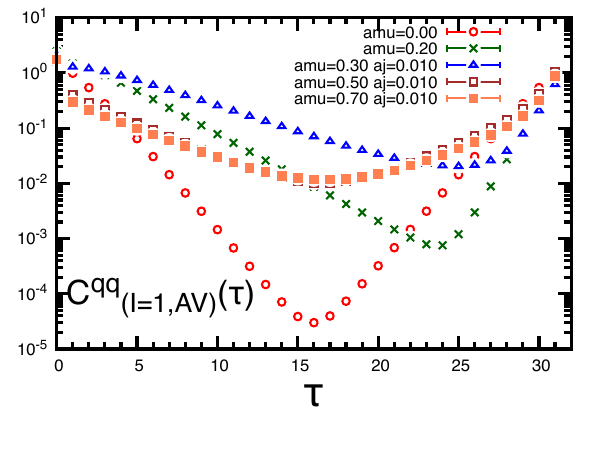}
\caption{The chemical potential dependence of the correlation function for the axial-vector diquark/antidiquark ($I=1, J^P=1^+$). \label{fig:corr-Diquark-iqn8}}
\end{figure}
As shown in Fig.~\ref{fig:corr-Diquark-iqn8}, at $\mu = 0$, the axial-vector diquark correlator is symmetric under the time-reversal transformation $\tau \leftrightarrow (N_\tau - \tau)$. As $\mu$ increases, this symmetry is gradually violated, but after the superfluid phase transition, the correlator tends to regain an approximately symmetric shape. This behavior is qualitatively the same as that observed in the iso-singlet scalar diquark channel discussed in the previous subsection.

\begin{figure}[htbp]
\centering
\includegraphics[width=.3\textwidth]{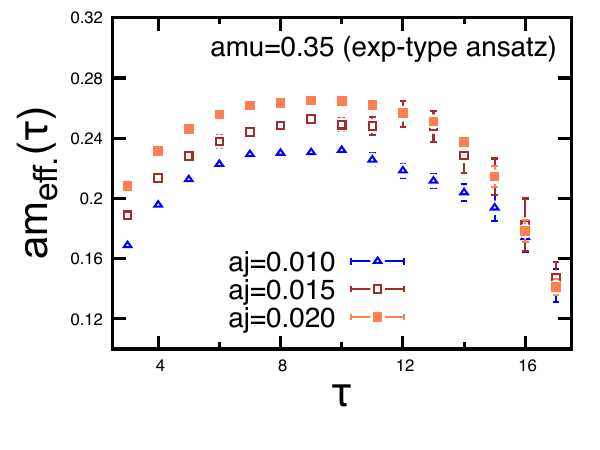}
\includegraphics[width=.3\textwidth]{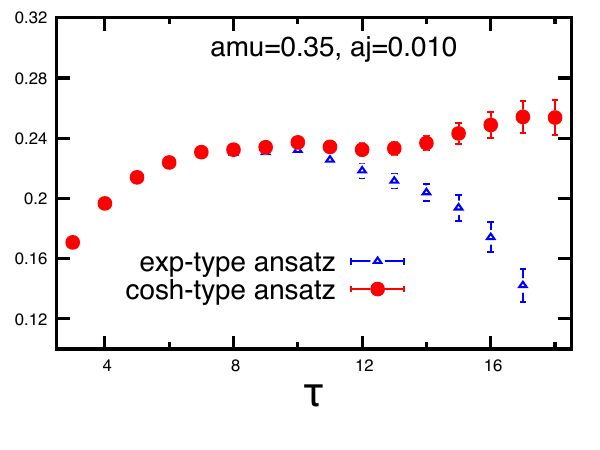}
\includegraphics[width=.3\textwidth]{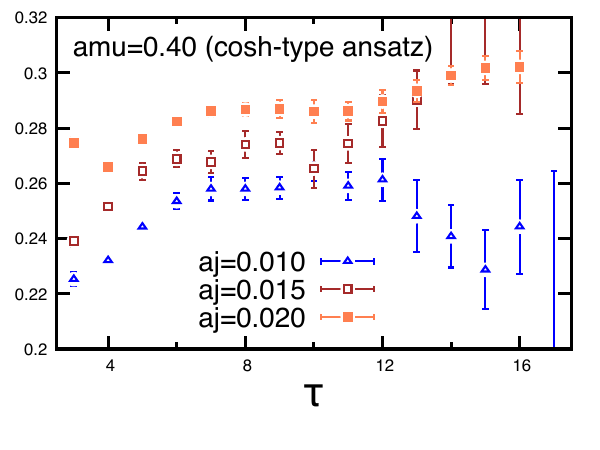}
\caption{The effective mass of the axial-vector diquark as a function of $\tau$. In the left panel, the effective masses obtained using the exp-type ansatz are plotted for various values of $j$ at $a\mu=0.35$.
The middle panel shows how the effective mass depends on the fitting ansatz, namely, the exp-type (Eq.~\eqref{eq:fit-fn-exp}) or the cosh-type (Eq.~\eqref{eq:fit-fn-cosh}), at ($a\mu=0.35,aj=0.010$). 
The right panel shows the $j$-dependence of the effective mass obtained using the cosh-type ansatz at $a\mu=0.40$. } \label{fig:eff-mass-tau-Diquark-iqn8}
\end{figure}

Figure~\ref{fig:eff-mass-tau-Diquark-iqn8} shows how the effective mass of the axial-vector diquark depends on $\tau$ for several values of $a\mu$. 
As in the previous subsection, we have extracted the effective mass at each $\tau$ using the exp-type ansatz in the low-$\mu$ region and the cosh-type ansatz in the high-$\mu$ region.
The left panel of Fig.~\ref{fig:eff-mass-tau-Diquark-iqn8} shows the effective masses at $a\mu = 0.35$ obtained using the exp-type ansatz for three values of $aj$. A plateau-like region can be seen around $\tau = 8$. In the middle panel, we compare the results obtained with the exp-type and cosh-type ansatzes at $a\mu = 0.35$ and $aj = 0.010$.  A similar plateau region appears for both ansatzes, yielding nearly identical plateau values.
For larger values of $\mu$, the plateau becomes more clearly visible when the cosh-type ansatz is employed. 
Indeed, the right panel shows the results at $a\mu = 0.40$, where a clear plateau region around $\tau = 8$ is observed for all the values of $aj$.  We therefore use the cosh-type ansatz for $a\mu \geq 0.40$.

\begin{figure}[htbp]
\centering
\includegraphics[width=.45\textwidth]{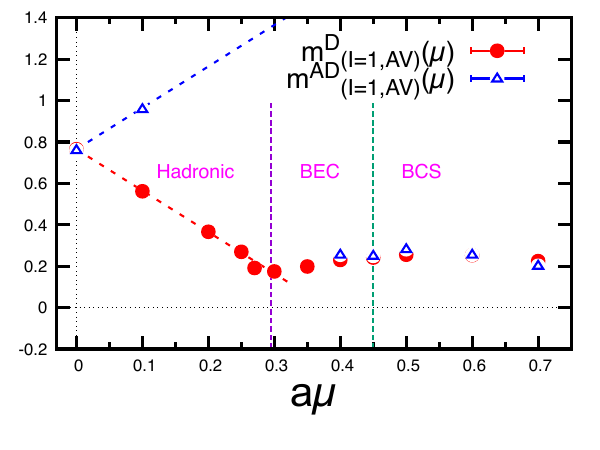}
\caption{The chemical potential dependence of the masses of the axial-vector diquark and antidiquark.
The dotted straight lines denote $m(\mu) = m(\mu=0) \pm 2\mu$, which are the theoretical predictions in Eq.~\eqref{eq:mass-shift}.
\label{fig:result-mass-diquark-8}}
\end{figure}

Finally, the results for the effective masses of the axial-vector diquark and antidiquark obtained after performing a linear extrapolation to the $j \to 0$ limit at each value of $a\mu$ are summarized in Fig.~\ref{fig:result-mass-diquark-8}. 
In the hadronic phase, the diquark mass exhibits an approximately linear dependence on $a\mu$, which is in agreement with the mass-shift prediction, Eq.~\eqref{eq:mass-shift}.
For the antidiquark mass, no reliable results are obtained in the range $0.20 \leq a\mu \leq 0.35$, because the number of available data points in the backward correlation functions is too limited to identify a stable plateau. For $a\mu \ge 0.40$, however, the resulting antidiquark mass becomes nearly equal to that of the diquark. This reflects the fact that the cosh-type ansatz provides a good description of the correlators in this region.

We now turn to the corresponding meson channel, namely, the $a_1$-meson channel. 
Here we consider the $\mu=i$ component of the axial-vector operators, namely, $\Gamma = i \gamma_5 \gamma_i$, evaluate the corresponding correlators, and take the zero spatial-momentum projection, $\vec{p} = \vec{0}$. That is why there is no need to consider a possible pion contribution related to the PCAC relation that arises from the non-conservation of the axial-vector current.

\begin{figure}[htbp]
\centering
\includegraphics[width=.45\textwidth]{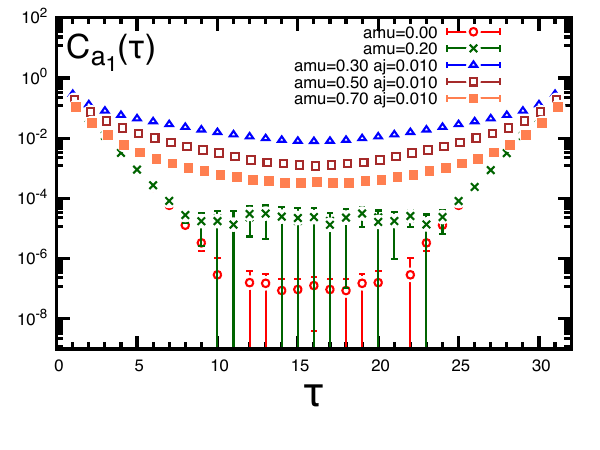}
\qquad
\includegraphics[width=.45\textwidth]{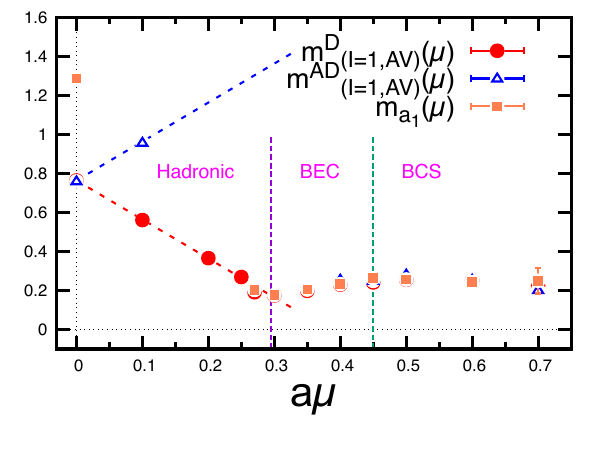}
\caption{(Left) The chemical potential dependence of the correlation function for the $a_1$-meson. 
(Right) The chemical potential dependence of the $a_1$-meson mass (orange-squared symbols).
For comparison, we also show the results for the axial-vector diquark and antidiquark masses presented in Fig.~\ref{fig:result-mass-diquark-8}. }\label{fig:corr-Meson-8}
\end{figure}

As in three-color QCD, this channel is very noisy in the hadronic phase in QC$_2$D. As shown in the left panel of Fig.~\ref{fig:corr-Meson-8}, however, the signal of $C_{a_1}(\tau)$ becomes much cleaner once the system enters the superfluid phase.
We compute the effective mass using a cosh-type ansatz for all values of $a\mu$ and extrapolate the results to the $j \to 0$ limit. The $\mu$-dependence of the resulting $a_1$-meson mass is summarized in the right panel of Fig.~\ref{fig:corr-Meson-8}, where we also plot the diquark and antidiquark results from Fig.~\ref{fig:result-mass-diquark-8} for comparison.
In the hadronic phase, we obtain a signal only at $\mu = 0$, where the effective mass is larger than $1.2$ in lattice units. For $a\mu < 0.27$, the correlator is too noisy to extract a reliable mass. Above $a\mu = 0.27$, the $a_1$-meson becomes degenerate with the diquark/antidiquark carrying the same quantum numbers. As in the previous subsection, this behavior can be interpreted as a consequence of mixing between the diquark and meson states in the superfluid phase, and indeed we observe the lightest state with $I=1, J^P=1^+$.

\subsection{Iso-singlet pseudoscalar meson and diquark}
\begin{figure}[htbp]
\centering
\includegraphics[width=.45\textwidth]{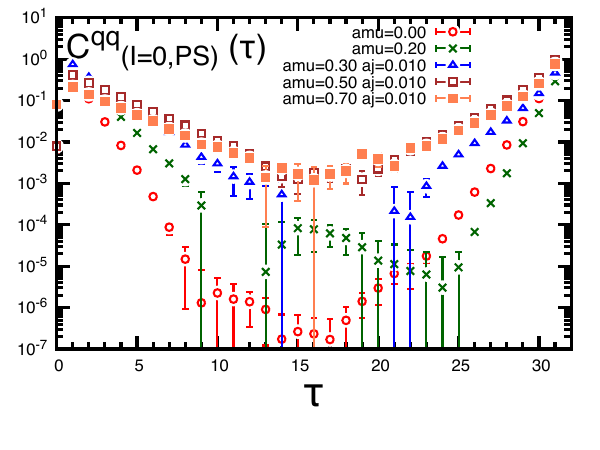}
\qquad
\includegraphics[width=.45\textwidth]{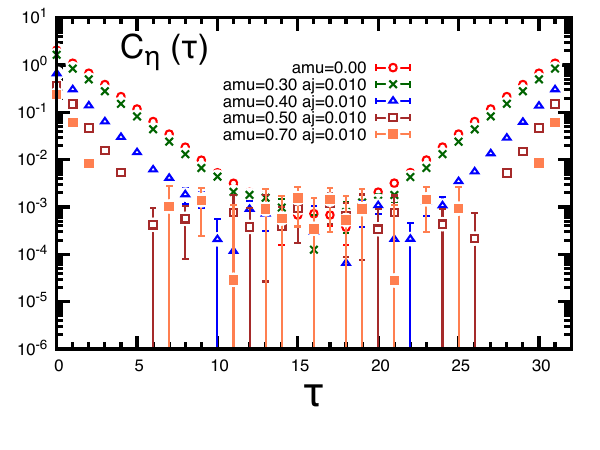}
\caption{The chemical potential dependence of the correlation functions for the PS diquark (left) and $\eta$-meson (right). \label{fig:corr-Diquark-3}}
\end{figure}

Let us now consider the iso-singlet PS channel. The correlation functions for the PS diquark and $\eta$-meson are shown in Fig.~\ref{fig:corr-Diquark-3}. 
In both correlators, we include the disconnected contributions. 
In the PS diquark correlator, however, the disconnected part is numerically negligible, since it is written in terms of anomalous propagators and is suppressed by powers of $j$, where $j$ is a small parameter. 
In contrast, the disconnected contribution in the $\eta$-meson correlator, which involves contractions of the normal propagator, is relatively large around $\tau= N_\tau/2$.
For the diquark correlator, the overall data are rather noisy at small $\mu$, and hence no reliable signal is obtained in the large-$\tau$ region. 
At larger $\mu$, the data become somewhat cleaner. 
For the meson correlator, on the other hand, the shape of the correlator can be identified in the hadronic phase at small $a\mu$, while for $a\mu \gtrsim 0.40$ it becomes so noisy that the signal is lost in the large $\tau$ range.

To extract the effective masses of the diquark and antidiquark, we use an exp-type ansatz for all the values of $\mu$ in this channel, since 
it is difficult to determine the location of the minimum ($N_\tau^{\mathrm{min}}$) required for a reliable cosh-type description. 
\begin{figure}[htbp]
\centering
\includegraphics[width=.45\textwidth]{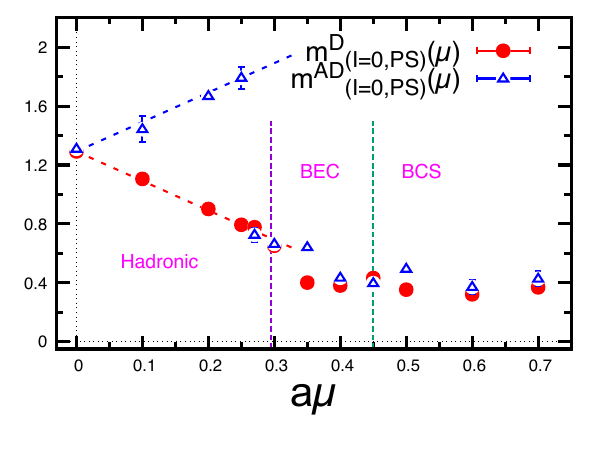}
\caption{The chemical potential dependence of the masses of the PS diquark and antidiquark.
The dotted straight lines denote $m(\mu) = m(\mu=0) \pm 2\mu$, which are the theoretical predictions in Eq.~\eqref{eq:mass-shift}.
\label{fig:result-mass-diquark-3}}
\end{figure}
The $\mu$-dependence of the resulting diquark and antidiquark masses is shown in Fig.~\ref{fig:result-mass-diquark-3}. In the hadronic phase, the masses follow the mass-shift prediction~\eqref{eq:mass-shift}.  Around $a\mu = 0.30$, the antidiquark becomes degenerate with the diquark, which again comes from the loss of baryon number conservation in the superfluid phase\footnote{At $a\mu=0.35$, the diquark and antidiquark masses appear to be split. Since the strength of baryon-antibaryon mixing generally depends on $\mu$, the higher mode may happen to be more visible at this point.}.

We next turn to the $\eta$-meson, which carries the same quantum numbers, $I=0, J^P=0^-$. As can be seen in Fig.~\ref{fig:corr-Diquark-3}, the signal of the $\eta$-meson is cleaner in the hadronic phase. 
\begin{figure}[htbp]
\centering
\includegraphics[width=.45\textwidth]{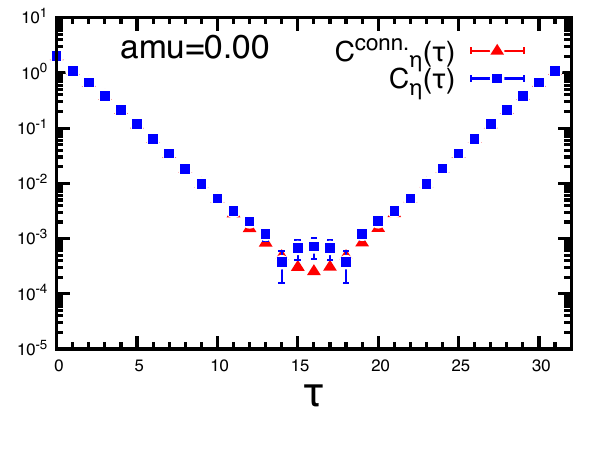}
\qquad
\includegraphics[width=.45\textwidth]{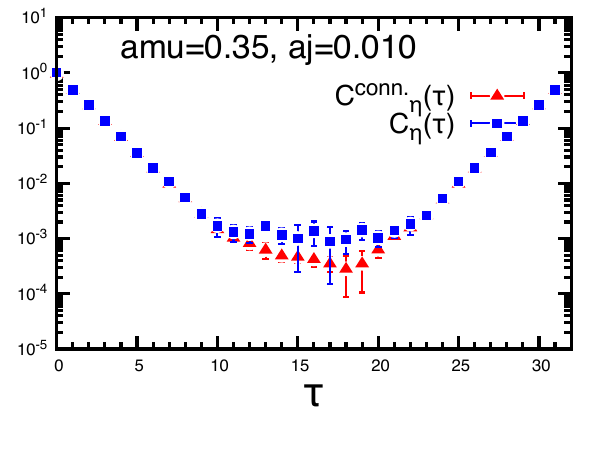}
\caption{Comparison between the total correlation function (blue-squared) and the connected part (red-triangle) for the $\eta$-meson in the hadronic ($a\mu=0.00$) and superfluid ($a\mu=0.35, aj=0.010$) phases. \label{fig:corr-Meson-3-mu000-035}}
\end{figure}
In Fig.~\ref{fig:corr-Meson-3-mu000-035}, we show the connected part of the correlator (red-triangle) and the total correlator including the disconnected contribution (blue-squared) at $a\mu = 0.00$ (left panel) and $0.35$ (right panel).
The disconnected part contributes at the level of order $10^{-2}$--$10^{-3}$ and hence becomes relevant in the long-$\tau$ region. 
As $\mu$ increases, the correlator falls off more rapidly with $\tau$, so that the effect of the disconnected contribution sets in even at smaller $\tau$, making the correlator increasingly noisy.
As a result, when we compute the effective mass at each $\tau$, the range over which a reliable signal can be extracted becomes progressively shorter. 
\begin{figure}[htbp]
\centering
\includegraphics[width=.45\textwidth]{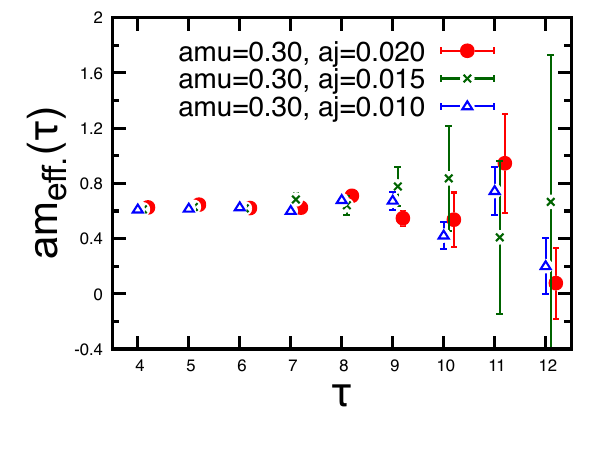}
\quad
\includegraphics[width=.45\textwidth]{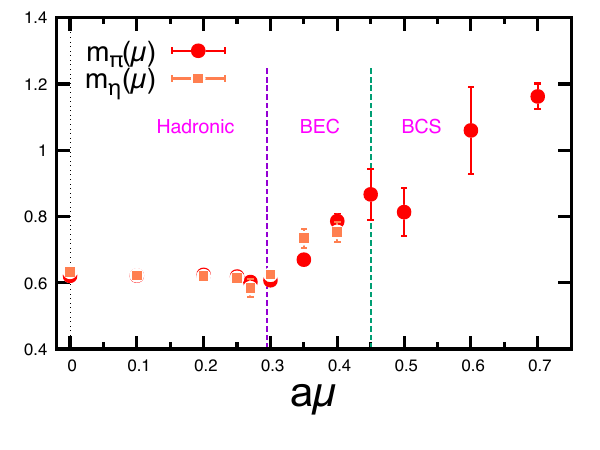}
\caption{(Left) The $\tau$-dependence of the effective mass of the $\eta$-meson at $a\mu=0.30$ with $aj=0.010,0.015$, and $0.020$. For clarity, the data points for $aj=0.015$ and $aj=0.020$ are slightly shifted along the horizontal axis.
(Right) The chemical potential dependence of the effective mass of the $\eta$-meson in the $j \rightarrow 0$ limit. No data are available for $a\mu \geq 0.45$ for the reasons discussed in the manuscript. For comparison, we also show the pion mass presented in Fig.~\ref{fig:mass-Meson-4-6}.
\label{fig:mass-Meson-3}}
\end{figure}
The left panel of Fig.~\ref{fig:mass-Meson-3} shows the $\tau$-dependence of the effective mass at $a\mu = 0.30$, where different symbols correspond to different values of $aj$. Around $\tau = 10$, the statistical error reaches roughly 50\%, and at even larger $\tau$, no clear plateau can be identified.
In the present study, we are practically able to extract the $\eta$-meson mass only up to $a\mu = 0.40$.

The $\mu$-dependence of the resulting masses is summarized in the right panel of Fig.~\ref{fig:mass-Meson-3}. 
We find that the $\eta$-meson becomes heavier once the system enters the superfluid phase.
According to analytical studies, it has been argued in both three-color and two-color QCD that instanton effects are suppressed at sufficiently high density~\cite{Schafer:1999fe, Kanazawa:2009ks}.
This implies an effective restoration of $U(1)_A$ symmetry and a reduction in the $\eta$-meson mass.
Although our data do not allow a definitive conclusion, the observed behavior does not show such a mass reduction, since the $\eta$-meson correlator exhibits a steeper falloff in $\tau$ in the high-$\mu$ regime, suggesting that the mass becomes heavier.
This is not inconsistent with our previous studies~\cite{Iida:2019rah, Iida:2024irv}, where we found that even in the BCS regime, the topological susceptibility remains almost unchanged from that in the hadronic phase, indicating that the instanton effects are not suppressed~\footnote{The naive connection between the reduction of  topological susceptibility and $U(1)_A$ symmetry restoration is still controversial. By means of the axial Ward-Takahashi identity, it was pointed that the topological susceptibility can be suppressed following the reduction of the chiral condensate, even when the instanton effect remains~\cite{Kawaguchi:2023olk,Fejos:2025oxi}. }.
Note, however, that as mentioned above, the $I=0$, $J^P=0^-$ channel involves meson–baryon mixing~\cite{Suenaga:2022uqn}; there remains a possibility that the measured mass does not correspond to the lowest one of this channel. A more detailed analysis in terms of mass eigenstates would be required for a proper interpretation.

As a technical point, the manner in which the mass increases is very similar to that observed in the iso-triplet PS (pion) channel  
as shown in the right panel of Fig.~\ref{fig:mass-Meson-3}. 
This may reflect the fact that it was hard to include the $\tau$ region where the disconnected contribution becomes dominant in this study. Our results suggest that an analysis with higher statistics to suppress the fluctuations of the disconnected contribution would be required for a more precise determination.

\begin{figure}[htbp]
\centering
\includegraphics[width=.45\textwidth]{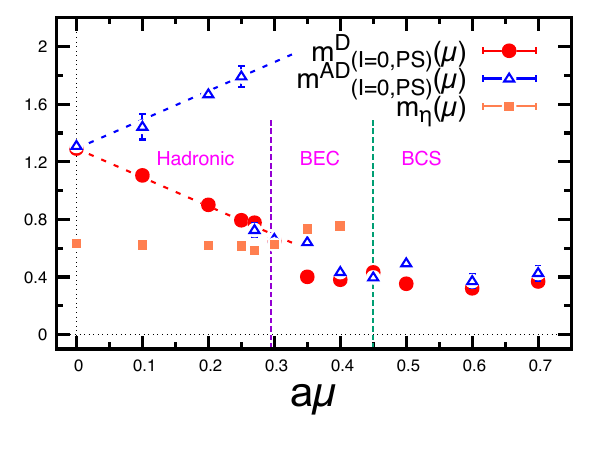}
\caption{The chemical potential dependence of the effective mass of the $\eta$-meson (orange-squared symbols).
For comparison, we also show the results for the PS diquark and antidiquark masses presented in Fig.~\ref{fig:result-mass-diquark-3}.
}\label{fig:result-mass-diquark-meson-3}
\end{figure}
Finally, in Fig.~\ref{fig:result-mass-diquark-meson-3} we compare the $\mu$-dependence of the effective masses of the hadrons in the iso-singlet PS channel, namely, the diquark, antidiquark, and $\eta$-meson. The $\eta$-meson is lighter in the hadronic phase, whereas in the superfluid phase the diquark and antidiquark become lighter. 
As in the previous subsection, a careful analysis allowing for meson–baryon mixing in the superfluid phase is left for future work.

\subsection{Iso-singlet vector meson and diquark}
\begin{figure}[htbp]
\centering
\includegraphics[width=.45\textwidth]{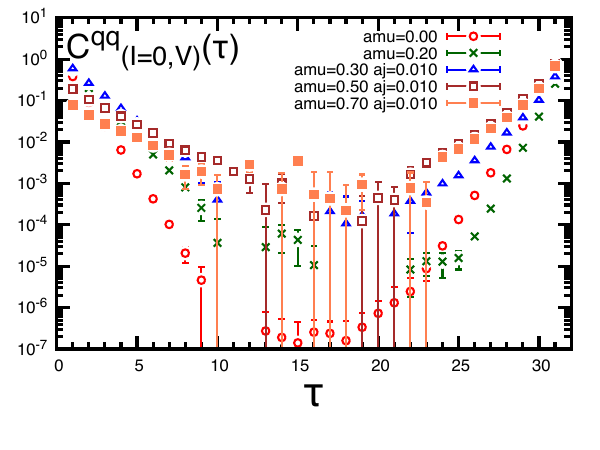}
\qquad
\includegraphics[width=.45\textwidth]{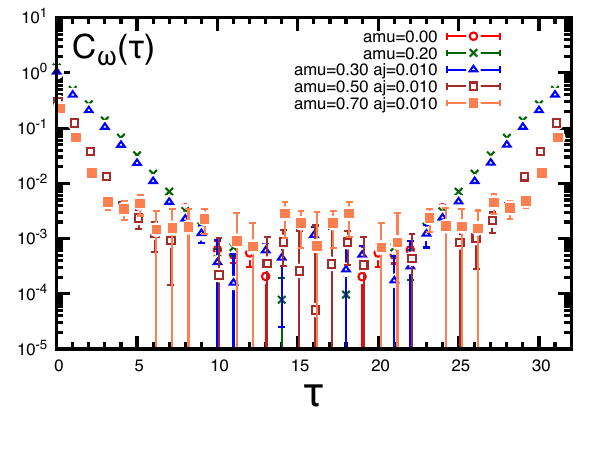}
\caption{The correlation functions for the vector diquark (left) and $\omega$-meson (right). \label{fig:corr-D-and-M-iqn5}}
\end{figure}

Next, we consider the iso-singlet vector channel, where we take the spatial component of the gamma matrix, $\Gamma = \gamma_i$, in the meson and diquark operators. The corresponding correlation functions are shown in Fig.~\ref{fig:corr-D-and-M-iqn5}. 
The vector-diquark correlator falls off more rapidly in the hadronic phase at small $\mu$, while in the superfluid phase at larger $\mu$ the slope becomes milder. In contrast, the meson channel, namely, the $\omega$-meson, exhibits a gentler slope in the hadronic phase, suggesting that it is lighter there.
In this channel, however, the signal-to-noise ratio is small compared with the other channels discussed in earlier subsections and hence no reliable data can be obtained in the large-$\tau$ region. 
Overall, the hadrons in this channel appear to be relatively heavy, which may indicate that they can couple to lighter modes.

For the extraction of the effective masses, we employ the exp-type ansatz for the diquark correlator and the cosh-type ansatz for the meson correlator, respectively, and then determine the masses from the plateau regions in both cases. 
\begin{figure}[htbp]
\centering
\includegraphics[width=.45\textwidth]{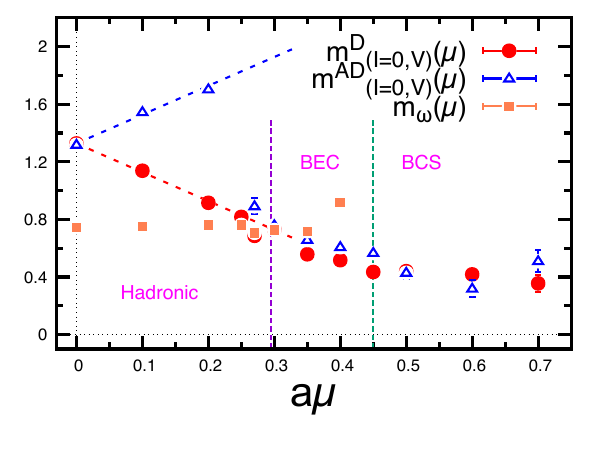}
\caption{The chemical potential dependence of the masses of the vector diquark, antidiquark, and $\omega$-meson.
The dotted straight lines denote $m(\mu) = m(\mu=0) \pm 2\mu$, which are the theoretical predictions in Eq.~\eqref{eq:mass-shift}.
\label{fig:result-mass-diquark-meson-5}}
\end{figure}
The resulting $\mu$-dependence of the masses is summarized in Fig.~\ref{fig:result-mass-diquark-meson-5}. As in the $\eta$-channel, we are able to reliably obtain the $\omega$-meson mass only up to $a\mu = 0.40$.
In the hadronic phase, the $\omega$-meson is lighter than the diquark and antidiquark. Around the superfluid transition point, the three states become nearly degenerate. At larger $\mu$, although the signal of the $\omega$-meson deteriorates, the diquark and antidiquark remain slightly lighter.

\subsection{Other noisy mesons}
\begin{figure}[htbp]
\centering
\includegraphics[width=.45\textwidth]{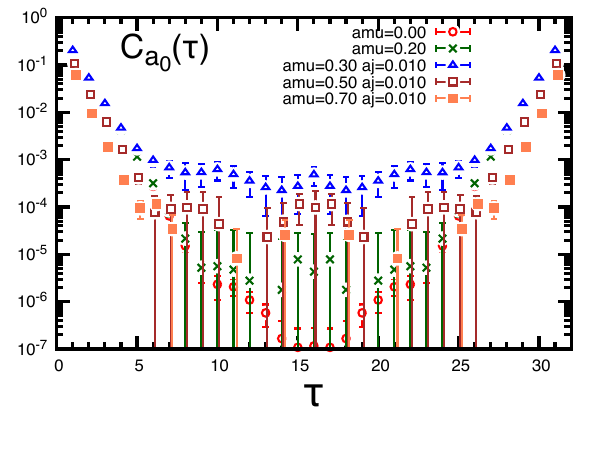}
\qquad
\includegraphics[width=.45\textwidth]{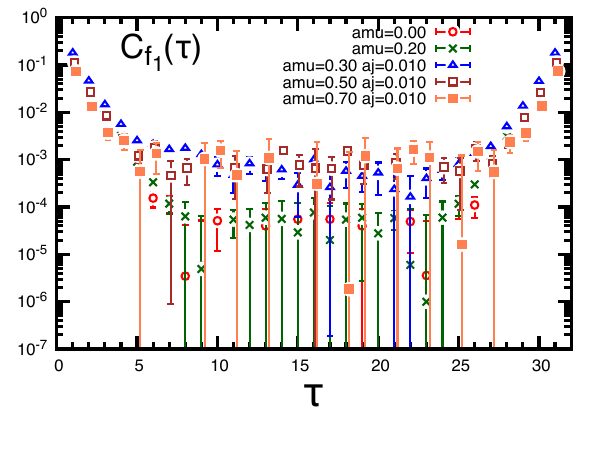}
\caption{Total correlation functions for the iso-triplet scalar ($a_0$) meson (left) and the iso-singlet axial-vector ($f_1$) meson (right). \label{fig:corr-Meson-2-and-7}}
\end{figure}

The remaining channels that can be considered are the iso-triplet scalar ($a_0$-meson) and the iso-singlet axial-vector meson ($f_1$-meson). The corresponding correlation functions are shown in Fig.~\ref{fig:corr-Meson-2-and-7}.
In both channels, the correlators are highly noisy in both the hadronic and superfluid phases.  Even so, the slopes of the correlators roughly suggest that the masses in these channels satisfy $am \gg 1$, and hence that these channels are especially heavy. In the present study, we do not attempt to determine the masses quantitatively, but restrict ourselves to presenting the qualitative behavior of the correlation functions.

\subsection{Summary of the $\mu$-dependence of the mass spectra}\label{sec:mass-summary}
We summarize the $\mu$-dependence of all the hadron masses obtained in this section.  So far, we have examined mesons and diquarks with the same quantum numbers in each subsection. Here, we focus mainly on the mass hierarchies among mesons and among diquarks.
\begin{figure}[htbp]
\centering
\includegraphics[width=.45\textwidth]{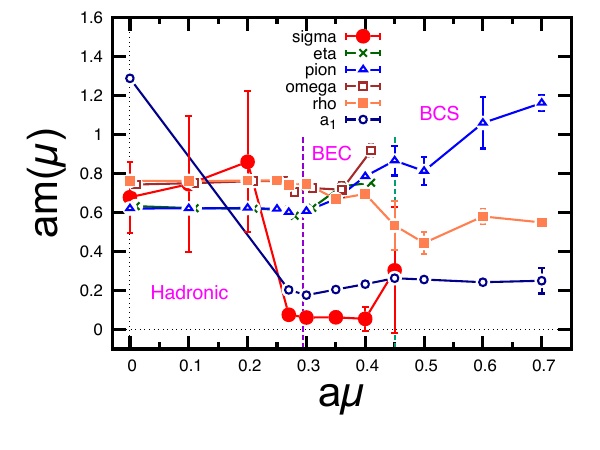}
\qquad
\includegraphics[width=.45\textwidth]{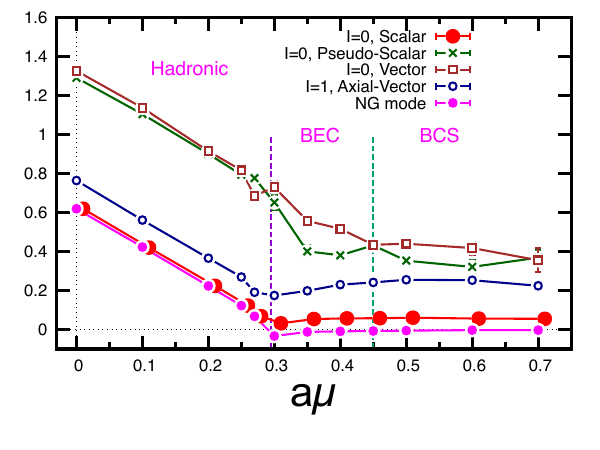}
\caption{The mass spectra for mesons (left) and for diquarks and the NG mode (right) \label{fig:summary-masses}}
\end{figure}
The left panel of Fig.~\ref{fig:summary-masses} collects the meson spectra, while the right panel shows the mass data for diquarks and the NG mode. 
As for the antidiquarks, they exhibit a linear scaling, $m(\mu) = m(\mu=0) + 2\mu$, in the hadronic phase, while most of them become degenerate with the corresponding diquarks with the same quantum numbers in the superfluid phase due to the loss of baryon-number conservation.  We, therefore, omit the results for the antidiquarks here.

As for the meson sector, the mass hierarchy in the hadronic phase as well as in the superfluid phase is given as follows:
\beq
\mathrm{Hadronic~phase:}&& m_\pi \lesssim m_{\eta}  < m_\sigma \mathrm{(noisy)} < m_\rho \sim m_\omega  \ll m_{a_1},\nonumber\\
\mathrm{Superfluid~phase:}&& m_\sigma \mathrm{(noisy)}  < m_{a_1} < m_\rho <  m_\pi \sim m_{\eta} \mathrm{(noisy)} \ll m_{\omega} \mathrm{(noisy)}. 
\eeq
The behavior in the hadronic phase shows a similar hierarchy to that of the meson sector in three-color QCD.
In the superfluid phase, the lightest hadron is the $I=0$ scalar meson, namely, the $\sigma$-meson. 
This can be understood as follows: The scalar diquark becomes the lightest as the NG boson associated with the spontaneous breaking of $U(1)_B$, and since the meson–baryon mixing is not resolved in the present analysis, we are effectively observing the lightest state with the same quantum numbers rather than a mass-eigenstate $\sigma$-meson. 
(In fact, in the presence of meson–baryon mixing, it may not be appropriate to use the term ``meson" itself.) Similarly, the lightness of the $a_1$ meson may come from meson–baryon mixing, since the diquark with the same quantum numbers, $I=1, J^P=1^+$, is observed as a light state.

As for the vector mesons, it is not clear from QCD sum rules for three-color QCD whether $m_\rho$ and $m_{\omega}$ become lighter 
in nuclear matter than at zero baryon density~\cite{Hatsuda:1991ez, Hatsuda:1995dy, Hatsuda:1996az, Klingl:1997kf, Rapp:1999ej, Zschocke:2002mn,Steinmueller:2006id}.
Our results for QC$_2$D indicate that the $\rho$-meson becomes slightly lighter in the superfluid phase than in the hadronic phase, while the $\omega$-meson becomes significantly noisier but seems to become heavier in the high-density region of the superfluid phase. 
The feature that $m_\rho$ tends to decrease in the superfluid phase is consistent with earlier lattice studies~\cite{Muroya:2002ry, Hands:2007uc} and appears to be robust.
As may be suggested from the fact that meson–baryon mixing, which is a characteristic feature of QC$_2$D, does not occur for the $\rho$ meson, the observed trend could persist also at nonzero baryon density in three-color QCD.
In contrast, the behavior of the $\omega$ meson may be affected by meson–baryon mixing, which is specific to QC$_2$D in the superfluid phase.

As for the PS mesons, $m_{\pi}$ increases approximately linearly with $\mu$ in the superfluid phase, which is roughly consistent with the ChPT prediction.
The $\eta$ meson also tends to become heavier once the system enters the superfluid phase.
From analytical studies~\cite{Schafer:1999fe, Kanazawa:2009ks}, it has been suggested in both three-color and two-color QCD that, at sufficiently high density, instanton effects are suppressed, leading to a reduction of the $\eta$ mass.
In contrast, our data rather suggest that the $\eta$ becomes heavier once the system enters the superfluid phase, although a definitive conclusion has yet to be drawn.
This does not necessarily contradict the above theoretical prediction. In our previous studies~\cite{Iida:2019rah, Iida:2024irv}, we found that instanton effects are not suppressed even in the high-density regime, as indicated by the topological susceptibility that remains nearly unchanged from that in the hadronic phase even in the BCS regime.
We again note that
the $I=0$, $J^P=0^-$ channel is subject to meson–baryon mixing. 
For a proper interpretation of this channel, a more careful analysis based on mass eigenstates would be necessary.

Let us now turn to the diquark sector. As shown in the right panel of Fig.~\ref{fig:summary-masses}, all the masses decrease in the hadronic phase according to the theoretical mass-shift prediction, Eq.~\eqref{eq:mass-shift}, while maintaining the same ordering as at $\mu=0$ even in the superfluid phase.
Thus, in both phases,
\beq
 m_{NG} \lesssim m_{I=0, S} < m_{I=1, AV} < m_{I=0, PS} \lesssim m_{I=0, V}. 
\eeq
The ordering at $\mu=0$ can be understood as follows: As long as the Pauli–G\"{u}rsey symmetry, which is exact in the massless limit, approximately remains, the $I=0$ scalar ($I=1$ axial-vector) diquark belongs to the same multiplet as the pion ($\rho$-meson).
This connection to the meson sector reflects the fact that the scalar and axial-vector diquarks are relatively light.

In the superfluid phase, the NG mode associated with $U(1)_B$ becomes almost massless when the $j \to 0$ limit is taken using the ChPT-motivated functional form of $j$, Eq.~\eqref{eq:j-sqrt}. 
At finite $j$, however, we observe that the NG boson mass exhibits a local minimum around $\mu = \mu_c$, as also predicted by ChPT.
We remark that the mass hierarchy among diquarks becomes less robust in the superfluid phase than in the hadronic phase.

\section{Chiral symmetry and correlation functions}\label{sec:result-chiral-sym}
In this section, we examine the restoration of chiral symmetry in the superfluid phase by comparing the correlation functions between chiral partners as explained in Sec.~\ref{sec:chiral-partner}.
As shown in Eqs.~\eqref{eq:S-P-relation} and \eqref{eq:V-A-relation}, it is expected that the correlators for the $I=0$ scalar meson ($\sigma$) and for the $I=1$ pseudoscalar meson ($\pi$), as well as those for the $I=1$ vector meson ($\rho$) and for the axial-vector meson ($a_1$), become degenerate at each spacetime point when the chiral symmetry is restored.

First, let us examine the $\rho$- and $a_1$-mesons.
\begin{figure}[htbp]
\centering
\includegraphics[width=.45\textwidth]{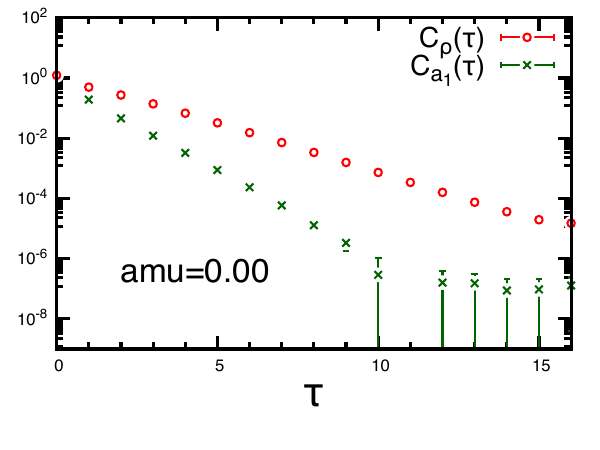}
\qquad
\includegraphics[width=.45\textwidth]{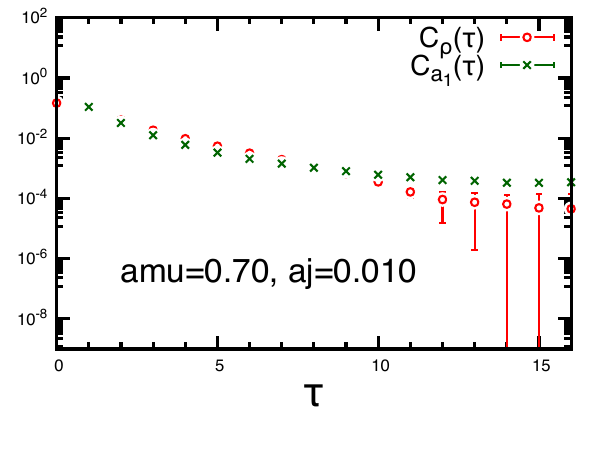}
\caption{Comparison between the correlation functions for chiral partners, the $\rho$ and $a_1$ mesons, in the hadronic phase ($a\mu=0.00$, left panel) and in the superfluid phase ($a\mu=0.70, aj=0.010$, right panel).  \label{fig:comp-corr-Meson-6-8}}
\end{figure}
In Fig.~\ref{fig:comp-corr-Meson-6-8}, we compare the correlation functions for the $\rho$-meson  (red-circle symbols) and the $a_1$-meson (green-cross symbols) at $a\mu = 0.00$ (left) and $a\mu = 0.70$ with $aj = 0.010$ (right). At $\mu = 0$, the two correlators exhibit clearly different slopes. In contrast, in the high-density region at $a\mu = 0.70$, the two correlation functions become remarkably similar, particularly for $\tau \lesssim 10$.

\begin{figure}[htbp]
\centering
\includegraphics[width=.45\textwidth]{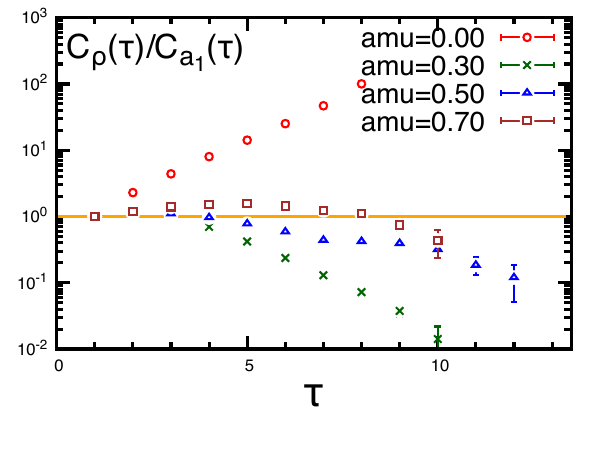}
\caption{The ratio of the correlation functions for chiral partners, the $\rho$ and $a_1$ mesons. The ratio of the correlators is normalized at $\tau = 1$.  \label{fig:ratio-corr-rho-a1}}
\end{figure}
To quantify these tendencies, we show the ratio $C_{\rho}(\tau)/C_{a_1}(\tau)$ in Fig.~\ref{fig:ratio-corr-rho-a1}.
Here, for $a\mu = 0.30, 0.50,$ and $0.70$, the data are obtained by first computing the correlators at several values of $aj$ at fixed $a\mu$ and then extrapolating the results to $j \to 0$ using a linear function.
The ratio $C_\rho(\tau)/C_{a_1}(\tau)$ is normalized to unity at $\tau = 1$ at each value of $a\mu$.
At $a\mu = 0.00$, the ratio is larger than unity because the $\rho$-meson is lighter than the $a_1$-meson. After entering the superfluid phase, however, as seen in Fig.~\ref{fig:summary-masses}, the $a_1$-meson becomes lighter than the $\rho$-meson. 
Consequently, for $a\mu \gtrsim 0.30$, the ratio becomes smaller than unity for most values of $\tau$. 
We find that the ratio tends to approach unity over a wider range of $\tau$ as $\mu$ increases.
As seen in Fig.~\ref{fig:ratio-corr-rho-a1}, this approximate degeneracy is observed up to $\tau \approx 8$ at the highest value of $a\mu=0.70$.
Note that the masses of the $\rho$ and $a_1$ are extracted from an even larger $\tau$ region, where the slopes of the correlators differ as shown in the right panel of Fig.~\ref{fig:comp-corr-Meson-6-8}.
As a result, the extracted masses of these two states remain different in Fig.~\ref{fig:summary-masses}.

We next examine the $\pi$- and $\sigma$-mesons.
As we have seen so far, the correlator for the $\sigma$-meson is rather noisy in the hadronic phase, while that for the $\pi$-meson is noisy in the superfluid phase.  Consequently, it is more difficult to draw clear conclusions about the correlator ratio between these chiral partners.
\begin{figure}[htbp]
\centering
\includegraphics[width=.45\textwidth]{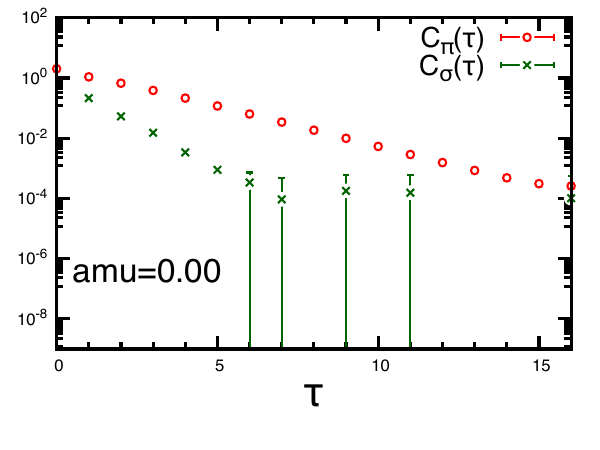}
\qquad
\includegraphics[width=.45\textwidth]{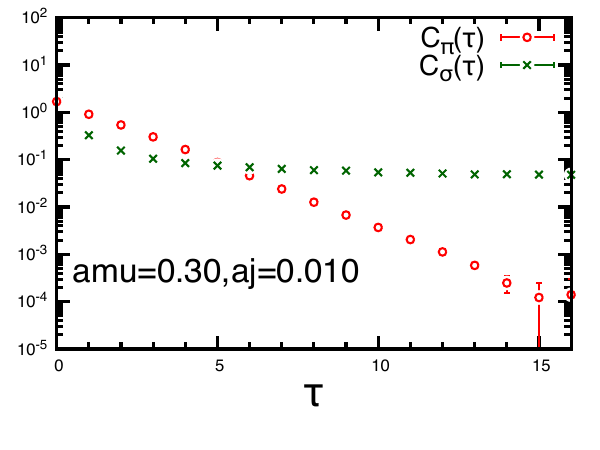}
\qquad
\includegraphics[width=.45\textwidth]{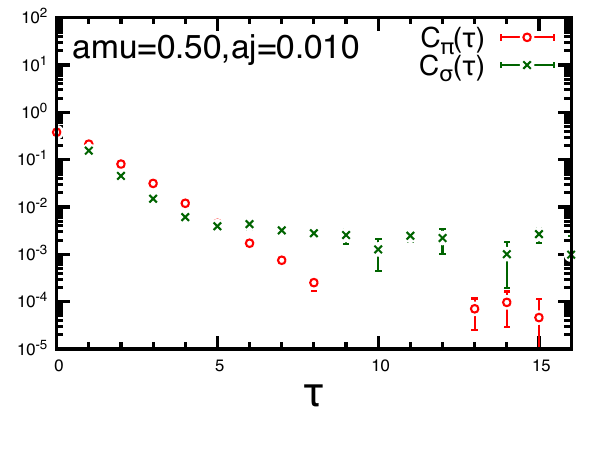}
\qquad
\includegraphics[width=.45\textwidth]{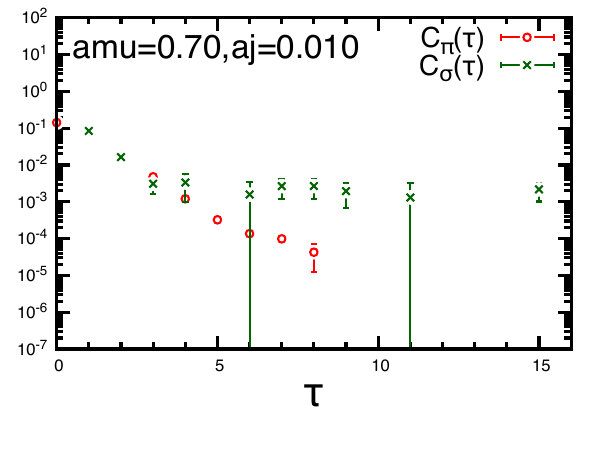}
\caption{Comparison between the correlation functions for chiral partners, the $\pi$- and $\sigma$-mesons, in the hadronic phase ($a\mu=0.00$), around the critical $\mu$ ($a\mu=0.30, aj=0.010$), around the BEC-BCS crossover point ($a\mu=0.50, aj=0.010$), and at the highest value of $a\mu$ ($a\mu=0.70, aj=0.010$). 
  \label{fig:comp-corr-Meson-4-2}}
\end{figure}
In Fig.~\ref{fig:comp-corr-Meson-4-2}, we compare the correlation functions for the $\pi$- and $\sigma$-mesons.
Here, we plot the results for $a\mu=0.00$, as well as for $a\mu=0.30$ with $aj=0.010$ (near the hadronic–superfluid phase transition), $a\mu=0.50$ with $aj=0.010$ (around the BEC–BCS crossover), and $a\mu=0.70$ with $aj=0.010$ (in the BCS phase).
At $a\mu=0.00$, the $\pi$-meson is obviously light compared with the $\sigma$-meson.
Then, the corresponding correlation functions do not agree at all even in a short-$\tau$ region.
After the system enters the superfluid phase, the $\sigma$-meson, which has the same quantum numbers as the NG boson, becomes lighter while the $\pi$ mass does not change significantly across the transition.  At $a\mu=0.30$, therefore, the correlators for the $\pi$-meson and for the $\sigma$-meson differ in a manner that is distinct from that at $a\mu=0.00$.
At $a\mu=0.50$ and $0.70$, the two correlators are found to be in better agreement although the signals are so noisy that only the region up to $\tau \lesssim 5$ is reliable.
We note that, around this regime, the chiral condensate also shows a plateau at a value smaller than that in the hadronic phase, as shown in Fig.~6 of Ref.~\cite{Iida:2024irv}, which implicitly suggests the restoration of chiral symmetry.

\begin{figure}[htbp]
\centering
\includegraphics[width=.45\textwidth]{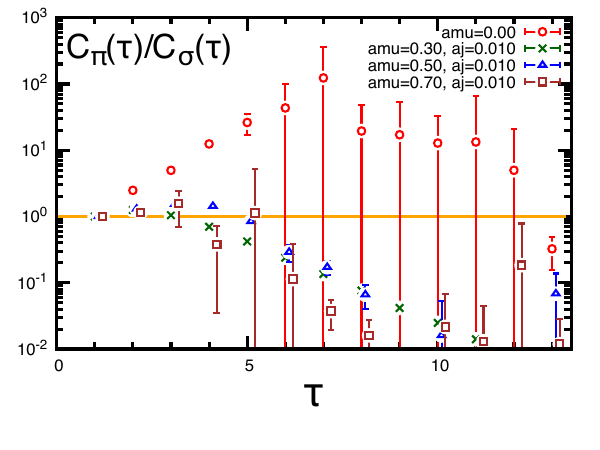}
\caption{The ratio of the correlation functions for chiral partners, the $\pi$- and $\sigma$-mesons. The ratio of the correlators is normalized at $\tau = 1$.  \label{fig:ratio-corr-pi-sigma}}
\end{figure}
Figure~\ref{fig:ratio-corr-pi-sigma} presents the ratio $C_{\pi}(\tau)/C_{\sigma}(\tau)$. 
If we take the $j \to 0$ extrapolation, some data for the $\pi$- and $\sigma$-meson correlators would become negative due to fluctuations.  To avoid this problem, here we plot the ratio at $j=0.010$ in the superfluid phase ($a\mu=0.30, 0.50,$ and $0.70$).
For all the values of $\mu$, the errors become so large for $\tau > 6$ that no definite conclusion can be drawn in this channel. By focusing on $\tau \lesssim 5$, however, one can see that the ratio tends to approach unity for larger $\mu$.

From these observations, we conclude that the ratio of the correlation functions for chiral partners exhibits a tendency toward unity at high density, which suggests partial restoration of chiral symmetry.

\section{Summary and future prospect}\label{sec:summary}
We investigate the chemical-potential dependence of the hadron spectra in two-color QCD using first-principles lattice simulations.
In this work, we compute two-point correlation functions for various hadrons by including, for the first time, the contributions from disconnected diagrams and extract the corresponding effective masses.

A detailed discussion of the resultant masses, including comparisons with analytical approaches such as ChPT, effective models, and QCD sum rules, has already been given in Sec.~\ref{sec:mass-summary}.
Here we repeat only the main results as follows:
In the meson sector, our results are summarized as
\beq
\mathrm{Hadronic~phase:}&& m_\pi \lesssim m_{\eta} < m_\sigma~\mathrm{(noisy)} < m_\rho \sim m_\omega \ll m_{a_1},\nonumber\\
\mathrm{Superfluid~phase:}&& m_\sigma~\mathrm{(noisy)} < m_{a_1} < m_\rho < m_\pi \sim m_{\eta}~\mathrm{(noisy)} \ll m_{\omega}~\mathrm{(noisy)}.
\eeq
In the diquark sector, the ordering 
\beq
m_{NG} \lesssim m_{I=0, S} < m_{I=1, AV} < m_{I=0, PS} \lesssim m_{I=0, V}
\eeq
persists in both phases; the NG mode associated with the spontaneous breaking of $U(1)_B$ is confirmed to be nearly massless in the superfluid phase.
We note that the mass hierarchy of the meson sector in the hadronic phase is similar to that in three-color QCD.

Furthermore, by comparing the correlators for chiral partners, we find indications of partial chiral-symmetry restoration at high density.

One of the remaining issues in the present analysis is that meson–baryon mixing in the superfluid phase has not been fully resolved.
Once the system enters the superfluid phase, therefore, not only the results for the diquark and antidiquark masses but also those for the meson masses tend to be nearly degenerate in some channels.
This implies that, among the states sharing the same quantum numbers $(I, J^P)$, only the lowest mass eigenstate is predominantly observed.
A more detailed analysis that explicitly disentangles these mixing effects by, e.g., solving a generalized eigenvalue problem (GEVP) would thus be highly desirable.
In addition, in the present study we have not discussed in detail the differences between the BEC and BCS regions within the superfluid phase.
In three-color QCD, the relation between the quark-hadron crossover and in-medium hadron properties has been discussed in Refs.~\cite{Schafer:1998ef,Abuki:2010jq}.
Motivated by these studies, it would be important to investigate how hadronic properties change across the superfluid phase of QC$_2$D in more detail.

From a more physical point of view, our results indicate that the quantum numbers of the lightest meson are changeable in the superfluid phase.
In the vacuum of ordinary QCD, the effective nucleon–nucleon potential is characterized by a repulsive core at short distances due to the Pauli blocking, an attractive pocket at intermediate distances of order 1 fm induced by $\rho$- and $\omega$-meson exchange, and a long-range part dominated by pion exchange.
Which meson dominates each distance region is closely related to the mass hierarchy of the mesons.
Our findings on the mass hierarchy in the meson sector suggest that this picture may be significantly altered at high density.

In the hadronic phase of QC$_2$D, the meson mass hierarchy is similar to that in three-color QCD.
Furthermore, first-principles calculations using the HAL QCD method have shown that the $\pi$–$\pi$ (and diquark–diquark) potentials also exhibit behavior similar to those in three-color QCD~\cite{Murakami:2023ejc}.
It would therefore be interesting to investigate hadron–hadron potentials in the superfluid phase of QC$_2$D, which might provide useful insights into the nature of hadronic interactions in dense three-color QCD.


\section*{Acknowledgment}
We are particularly grateful to T.~Hatsuda for valuable suggestions.
Discussions at the ``Workshop on Heavy-Quark Exotics'' were very helpful in completing this work.
E.~I.~would like to thank Y.~Hidaka and S.~Reddy for useful comments.
K.~I.\ is supported by JSPS KAKENHI with Grant No.\ 23K25864. 
K.~I.\ and E.~I.\ are supported by JSPS KAKENHI with Grant Number 25K01001. 
The work of E.~I.\ is supported by 
JST SQAI with Grant Number JPMJPF2221,  
JST CREST with Grant Number JPMJCR24I3.  
E.~I.\ and D.~S.\ are supported by JSPS KAKENHI with Grant Number 23H05439. 
D.~S.\ is also supported by JSPS KAKENHI No.~23K03377, No.~23H05439 and No.~25K17386.
K.~M.\ is supported in part by Grants-in-Aid for JSPS Fellows (Nos.\ JP22J14889, JP22KJ1870) and by JSPS KAKENHI with Grant No.\ 22H04917. 
This work is also supported by Program for Promoting Researches on the Supercomputer ``Fugaku'' (Simulation for basic science: from fundamental laws of particles to creation of nuclei) and (Simulation for basic science: approaching the new quantum era), and Joint Institute for Computational Fundamental Science (JICFuS), Grant Number JPMXP1020230411. 
This work is supported by Center for Gravitational Physics and Quantum Information (CGPQI) at Yukawa Institute for Theoretical Physics.
The authors used ChatGPT (OpenAI) solely for English-language copy editing of author-written text, including grammar, readability, and style (and this sentence). The authors reviewed all edits and take full responsibility for the final manuscript.


\bibliographystyle{JHEP}
\bibliography{2color}

\end{document}